*Research article*

# An overlay multicast routing method based on network situational awareness and hierarchical multi-agent reinforcement learning

**Miao Ye[1], Yanye Chen[1], Yong Wang[2], Cheng Zhu[1,3], Qiuxiang Jiang[4,*], Gai Huang[1] and Feng Ding[1]**

[1] School of Information and Communication, Guilin University of Electronic Technology, Guilin 541000, China

[2] School of Computer Science and Information Security, Guilin University of Electronic Technology, Guilin 541000, China

[3] Information Center, Guilin Medical University, Guilin 541000, China

[4] School of Optoelectronic Engineering, Guilin University of Electronic Technology, Guilin 541000, China

* **Correspondence:** Email: jiangqiuxiang@guet.edu.cn.

**Abstract:** Compared with IP multicast, Overlay Multicast (OM) trees constructed at the application layer offer superior compatibility and flexible deployment advantages in heterogeneous, cross-domain networks. However, OM implementations under traditional network architectures suffer from weak adaptability to highly dynamic traffic due to their lack of awareness of underlying physical resource states. Moreover, reinforcement learning-based approaches fail to decouple the multi-objective tightly coupled nature of OM, resulting in high computational complexity, slow policy convergence, and insufficient stability. To address these challenges, we proposed a MA-DHRL-OM routing method. First, leveraging the centralized topological view provided by Software-Defined Networking (SDN), the method collected link-state information and constructed a traffic-aware feature model to provide multi-dimensional decision support for OM path planning. Second, within a unified framework that integrates multi-agent reinforcement learning and hierarchical reinforcement learning, MA-DHRL-OM solves for the optimal OM tree as follows: The hierarchical learning architecture decomposes the construction of the OM tree into a two-stage subtask framework. By designing tailored decision logic and reward signal feedback mechanisms for upper- and lower-layer agents, it achieved hierarchical decoupling of the high-dimensional OM problem, effectively reducing the action space dimensionality and enhancing policy convergence stability. Moreover, the multi-agent collaboration mechanism enabled each agent to make independent decisions based on its local observations, thereby balancing multi-objective optimization while improving the algorithm's overall scalability and adaptability. Extensive simulation experiments demonstrated that, compared with existing methods, MA-DHRL-OM achieves superior performance in optimizing key metrics such as delay, bandwidth utilization, and packet loss rate while exhibiting more stable convergence behavior and greater flexibility in OM routing decisions.

**Keywords:** overlay multicast; software-defined networking; hierarchical reinforcement learning; multi-agent reinforcement learning; multicast routing

## 1. Introduction

With the rapid and vigorous development of modern network communication technologies and application demands, network transmission data volumes continue to grow at an explosive rate. In fields such as multimedia communication, video distribution, and data center interactions, the need for efficient and reliable data distribution mechanisms is constantly increasing [1]. Among these, IP multicast technology, as a crucial means to optimize network exploitation and enhance transmission efficiency, can significantly improve the overall operational efficiency and performance of communication systems [2]. However, IP multicast technology faces limitations and shortcomings during deployment and application: First, implementing IP multicast functionality relies on routers in the underlying network to maintain and update forwarding state for each multicast session [3,4]. This stateful forwarding mechanism deviates from the "stateless" design principle consistently followed at the IP layer, increasing protocol processing and state maintenance complexity; second, IP multicast lacks fine-grained access control and identity authentication mechanisms for member access, making reliability and security implementation more complex. Furthermore, cross-domain IP multicast relies on inter-domain routing information exchange, typically confining its application to single autonomous systems or local networks [5]. More critically, factors such as rapidly changing network state information and dynamic group membership further increase the management complexity and deployment costs of IP multicast, limiting its adoption and application in cross-domain scenarios [6].

Overlay Multicast (OM) serves as an alternative data distribution mechanism to IP multicast [7]. It aims to achieve efficient, scalable multipoint data distribution by constructing a virtual multicast forwarding structure at the application layer without relying on underlying network devices supporting multicast protocols. This mechanism shifts multicast functionality from the network layer to the application layer. It exploits logical connections between end hosts or dedicated servers to construct a multicast forwarding tree operating atop the overlay network. Through this approach, OM effectively circumvents the dependency on deploying IP multicast protocols within the underlying network infrastructure [8]. Within the overlay multicast architecture, terminal nodes not only perform basic data reception but also possess relaying capabilities. They can redistribute received multicast data to other terminal nodes [9]. Unlike traditional IP multicast, which relies on underlying network devices for data replication and distribution, overlay multicast employs application-layer virtual multicast topologies for multicast planning. This enables nodes to function as data receivers and, when necessary, as forwarding nodes [10]. Leveraging this mechanism, OM is commonly employed for efficient data distribution across domains or multi-operator environments, such as distributed video conferencing, content delivery networks (CDNs), and multimedia streaming [11].

Although overlay multicast effectively circumvents limitations of IP multicast, such as cross-domain deployment, protocol compatibility, and underlying device dependencies, by constructing virtual multipoint communication topologies at the application layer, building efficient, stable, and scalable overlay multicast structures remains challenging in highly dynamic network traffic environments [12]. First, constructing the optimal overlay multicast tree is a classic NP-hard problem [13]. Its computational complexity increases rapidly with network scale, making direct solution impractical. Second, overlay multicast nodes rely solely on application-layer logic to build virtual topologies, lacking direct access to underlying network state information such as link bandwidth, latency, and packet loss rate. This hinders timely responses to rapid network environment changes [14]. This lack of awareness makes it difficult for OM to dynamically adjust data forwarding paths when facing link fluctuations, node load changes, and burst traffic. This can lead to congestion, increased latency, and higher packet loss rates, thereby affecting overall transmission quality and user experience. Therefore, constructing an optimal overlay multicast spanning tree necessitates precise perception of underlying network information combined with efficient algorithm. This enables rapid scheduling and continuous optimization in dynamic environments, significantly enhancing scalability and robustness in large-scale, complex dynamic networks.

Traditional network architectures rely on local information exchange among distributed nodes to construct forwarding topologies. They lack timely access to global information such as network link quality, node load, and bandwidth resources [15]. This results in suboptimal forwarding path selection and may cause high latency, congestion, or degraded quality of service. Compared to traditional architectures, Software-Defined Networking (SDN) technology addresses this critical shortcoming in traditional architectures by centralizing the control plane [16]. This empowers the controller to receive a comprehensive view of the entire network topology, link status, and resource utilization from the data plane, thereby compensating for these limitations in OM applications [17]. Leveraging the real-time network state awareness capabilities of SDN controllers enables accurate assessment of communication overhead and transmission performance between nodes [18]. This provides more targeted selection criteria for determining optimal multicast data forwarding paths, laying a solid foundation for building smarter, more efficient, and scalable overlay multicast systems [19].

For NP-hard combinatorial optimization problems in overlay multicast, research has proposed heuristic algorithms based on greedy search policies (e.g., H-MDM [20] and HMTC [21]), as well as swarm-based optimization algorithms (e.g., genetic algorithms [22] and ant colony optimization ACO [23]). These methods typically fail to adapt to rapidly changing network state information, being unable to promptly adjust the connection structure and routing paths of the overlay multicast tree. Greedy policies based on shortest paths or minimum delay may become ineffective during link congestion. As network scale expands, interactions between nodes grow increasingly complex, causing the network state space to expand exponentially. Traditional heuristic algorithms, which make decisions based on fixed models or rules, struggle to effectively process high-dimensional, unstructured state information and are prone to local optimum. While swarm intelligence algorithms possess strong global search capabilities, they suffer from high computational complexity, slow convergence, and insufficient adaptability to dynamic networks.

Deep reinforcement learning algorithms, as a data-driven intelligent decision-making method, can automatically extract key features from massive high-dimensional network state information and identify potential behavioral patterns [24,25]. By continuously learning and adjusting policies through interaction with the environment, it is highly suited for dynamically changing network environments, achieving global performance optimization under uncertainty and multi-objective constraints [26,27]. Current approaches include applying deep reinforcement learning to multicast routing problems, such as MOMR-NGO [28] and TABDeep [29]. However, applying reinforcement learning to overlay multicast tree problems presents challenges. Since overlay multicast operates at the application layer of end nodes, path construction relies on logical connections between them. Moreover, designing reinforcement learning action policies must simultaneously consider relay node selection and forwarding path construction, leading to sparse reward issues. The high dimensionality of the problem and the increased dimension of the action space in reinforcement learning lead to slow training convergence, inflexible policy, and difficulty in meeting real-time adjustment requirements. Hierarchical reinforcement learning can structurally decompose complex tasks into manageable subtasks [30]. Applying hierarchical reinforcement learning to OM effectively reduces the problem's state and action dimensions while aligning the decision structure with the inherent characteristics of OM [31]. However, when applying hierarchical reinforcement learning to solve OM, lower layers often need to solve independent problems from the overlay region to designated destination nodes. Moreover, direct computation may delay response times for upper-layer agents. Multi-agent reinforcement learning effectively addresses this issue. Multi-agent reinforcement learning distributes subtasks among collaborating agents [32]. Each agent independently plans and makes decisions based on its local perspective, achieving coordinated multi-objective optimization. This approach distributes the time complexity of lower-layer problems and enhances operational efficiency [33].

By analyzing methods, we propose a multi-agent deep hierarchical reinforcement learning-based OM (MA-DHRL-OM), an intelligent solution based on multi-agent hierarchical reinforcement learning for OM routing under SDN. By decomposing the overlay multicast tree construction problem into a two-stage task, it designs deep reinforcement learning algorithms for upper and lower layers. For the second stage, optimizing forwarding node selection and route construction based on lower-layer network topology, a multi-agent collaborative solution mechanism is devised. Within this framework, each routing agent corresponding to a destination node operates logically independently, possessing distinct state and action spaces. Agents make autonomous decisions based on their environmental perception and policy network. This decentralized design not only enhances the algorithm's scalability and flexibility but also effectively supports parallel computation. Consequently, it enables more efficient exploration and optimization of multi-path multicast routing schemes, thereby elevating the overall intelligence and execution efficiency of OM tree routing construction.

The major contributions of this paper are as follows:

1) To address the issues of traditional overlay multicast lacking a global network view, inflexible routing path adjustments, and poor dynamic adaptability, we propose an SDN-based OM architecture: MA-DHRL-OM. Leveraging the centralized control feature of SDN controllers, we collect key performance indicators such as link bandwidth, delay, and packet loss rate from the underlying physical network to construct network state representations. This provides topological information support for reinforcement learning-driven overlay multicast strategies, effectively enhancing the response capability of overlay multicast tree routing strategies to dynamic changes in the network environment and improving the adjustability of path optimization strategies.
2) For the optimal construction of overlay multicast that requires simultaneous consideration of node selection and path planning, leading to high coupling between paths of the overlay multicast tree and making it difficult to solve due to the high dimensionality of the problem, we design a collaborative decision-making mechanism between upper layer and lower layer strategies. This includes devising a multi-dimensional network state representation method based on information such as link bandwidth, transmission delay, packet loss rate, and network topology. Upper-layer agents make decisions on destination node sequences according to the correlation of global node topologies, while lower-layer agents perform optimal source node selection and route construction under link constraints. Through task division and delegation, upper layer and lower layer agents achieve joint optimization of overlay multicast. Considering the differences in characteristics of tasks at different layers, hierarchical action spaces and reward functions are designed to enhance the convergence stability and adaptability of the strategy.

3) To tackle the high computational complexity and tendency to fall into local optima faced by overlay multicast routing methods in large-scale networks, we propose a multi-agent distributed routing construction mechanism. This mechanism decomposes the overall routing construction task into independent subtasks within a hierarchical architecture, where each agent is responsible for routing path planning for a single destination node, independently completing source node selection and route construction. The visitation sequence of destination nodes generated by upper-layer strategies provides global coordination constraints for individual agents, ensuring consistency in independent decision outcomes. This architecture enables parallel processing of routing computation tasks, significantly reduces algorithmic time complexity, enhances system scalability, and avoids the issue of exponentially growing decision complexity found in centralized approaches.

The remainder of this paper is organized as follows: In Section 2, we present the related work. In Section 3, we formulate the optimal overlay multicast problem. In Section 4, we introduce the SDN-based MA-DHRL-OM architecture for OM. In Section 5, we detail the proposed algorithm. In Section 6, we describe the experimental setup and present the performance evaluation results. In Section 7, we conclude the paper and discuss future work.

## 2. Related work

In this section, we primarily review approaches for solving overlay multicast and multicast scenarios, including traditional algorithms, swarm-based optimization algorithms, and artificial intelligence algorithms.

Traditional approaches typically rely on heuristic rules, structural optimization, or predefined mechanisms for path construction and resource allocation, lacking adaptive learning capabilities in the decision-making process. CAO et al. [20] proposed a heuristic-like approximation algorithm for the degree-constrained minimum-delay spanning tree problem, which employs a Dijkstra-like method to compute the longest-delay path from each node to the source and dynamically checks the parent node's degree during tree construction; if the degree constraint is violated, the algorithm backtracks to a suboptimal path, significantly reducing end-to-end delay. Wang et al. [34] addressed the source-side bottleneck and resource inefficiency in cloud networks caused by the absence of underlying IP multicast by introducing a greedy breadth-first tree construction algorithm with a maximum fanout constraint; this algorithm prioritizes placing nodes closer to the source at lower levels and expands only across switches or subnets when the remaining uncovered demand exceeds the node's capacity. Ying et al. [21] tackled bandwidth estimation distortion due to shared bottleneck links by proposing HMTC, a centralized heuristic algorithm that first constructs an initial tree using conventional independent bandwidth estimates and then iteratively removes the edge with the minimum bandwidth, seeking a replacement edge that reconnects the resulting subtrees and improves overall bandwidth, performing local refinements until no further improvement is possible. These traditional algorithms exhibit low computational complexity and are easy to implement, making them suitable for distributed environments; however, they generally rely on greedy strategies or other local search mechanisms, which are prone to local optima, target relatively narrow optimization objectives, and demonstrate limited adaptability to network topology changes or variations in demand parameters, often leading to network congestion in complex scenarios.

Intelligent optimization methods leverage path planning and structural optimization to exhibit strong global search capabilities and favorable parallelism when addressing complex problems. By enabling cooperative interactions among multiple individuals, these methods explore promising solutions across the global solution space, offering robustness and adaptability. Tseng et al. [35] proposed NGA, a genetic algorithm that directly manipulates tree structures to construct minimum-cost multicast trees that satisfy delay and degree constraints. The tree serves as the chromosome, and an enhanced Prim-based strategy is integrated into initialization, crossover, and mutation operations. Edge weights are heuristically designed to balance cost and feasibility, while constraint-violating individuals are penalized to guide convergence toward feasible, low-cost solutions. Lin et al. [36] formulated the multicast routing problem as a degree- and delay-constrained minimum-cost overlay spanning tree problem and introduced DDMOCST-GA, a heuristic genetic algorithm. This algorithm employs edge-set encoding and uses an improved Prim's algorithm to generate a feasible initial population. During crossover, common edges from both parents are inherited first, followed by greedy connection of remaining components using low-cost edges; mutation replaces high-cost edges with randomly selected lower-cost alternatives. Liu et al. [37] addressed the forwarding limitation of end-hosts in application-layer multicast by proposing a gene-pool-based genetic algorithm for optimal multicast tree construction. The method minimizes total transmission delay and utilizes a gene-pool mechanism to merge parental trees while eliminating cycles. Li et al. [38] tackled node congestion and premature convergence in single-session application-layer multicast by introducing a discrete artificial fish swarm algorithm (DAFSA). Individuals are encoded as Steiner node matrices, and feasible routing trees respecting degree constraints are constructed via an enhanced Prim's algorithm. Tseng et al. [23] developed ADD-CMST, an improved ant colony algorithm for the delay- and degree-constrained minimum-cost tree problem in application-layer multicast. It selects edges using an enhanced Prim procedure and a multi-dimensional heuristic function, prevents premature convergence through combined local and global pheromone updates, and enforces constraints via an exponential penalty function. Although these swarm intelligence-based optimization algorithms demonstrate strong global exploration and avoid local optima, they suffer from slow convergence and high computational overhead due to population-based iteration and heuristic construction mechanisms, rendering them inefficient and ill-suited for dynamic network environments.

Machine learning algorithms demonstrate tremendous potential in optimizing link load balancing and system traffic scheduling within multicast networks. Jihun et al. [39] addressed the minimum-cost multicast tree problem in SDN by formulating tree construction as a Markov decision process and proposed a meta-reinforcement learning algorithm. Their approach employs a model-based meta-RL framework, where an A3C-LSTM meta-agent learns an optimal link selection policy. Li et al. [28] tackled high energy consumption and latency in large-scale multicasts by introducing MOMR-NGO, a deep Q-network (DQN)-based algorithm that initializes DQN weights using the Northern Goshawk Optimization algorithm. Their state representation incorporates link-sharing degree to promote traffic aggregation, and optimal trade-off solutions are obtained via a reward function combined with a Pareto-front selection mechanism. Xia et al. [29] proposed TABDeep, a deep reinforcement learning algorithm for distributed subtree scheduling in online multicast within elastic optical networks. By jointly optimizing destination grouping, source selection, subtree construction, and scheduling order, TABDeep enables efficient real-time handling of dynamic multicast requests. Ye et al. [40] addressed inter-domain multicast routing in multi-domain software-defined wireless networks (SDWNs) with MA-CDMR, a multi-agent deep reinforcement learning framework. The problem is decomposed into inter-domain and intra-domain sub-problems, each managed by a dedicated agent; decentralized learning and hybrid training enable direct acquisition of optimal routing policies. These reinforcement learning algorithms directly optimize policies with respect to the objective, eliminating the need for handcrafted heuristic rules and offering strong dynamic adaptability and real-time responsiveness, making them well-suited for scenarios with frequently changing network states. By integrating techniques such as graph neural networks and multi-agent collaboration, they effectively model complex network topologies and distributed decision-making processes. Once trained, their high inference efficiency overcomes the computational inefficiency inherent in swarm intelligence methods. Through self-learning policy adaptation, these approaches successfully address the limitations of traditional multicast algorithms in complex network environments. However, they fail to adequately account for the hierarchical nature of global planning and local decision-making in overlay multicast tasks. The absence of effective task decomposition and modular design leads to poor adaptability and suboptimal coordination efficiency in multi-source, multi-destination networks, resulting in high computational costs and unstable training dynamics.

OM solutions often treat OM construction as a monolithic decision problem without explicit hierarchical decomposition or coordination mechanisms. As a result, training may suffer from non-stationarity and policy interference, leading to slow or unstable convergence and susceptibility to suboptimal solutions. To address these issues, we propose MA-DHRL-OM, an SDN-enabled overlay multicast routing framework that couples multi-agent learning with deep hierarchical reinforcement learning. The OM tree construction is formulated as a two-stage process: The upper layer generates the destination ordering, while the lower layer constructs routes by selecting parent/forwarding paths conditioned on that ordering. At the lower layer, dedicated agents are assigned to destination-specific routing subproblems to enable coordinated optimization. In addition, a hybrid online–offline training scheme is adopted to reduce dependence on real-time interaction and improve convergence efficiency.

## 3. Problem statement and modeling

### *3.1. Problem description*

Overlay multicast is a communication method that transmits data from a source node to multiple destination nodes at the application layer of a computer. Unlike traditional IP multicasting, overlay multicast relies on the collaboration between terminal nodes to construct a tree-like topology through logical links, thereby implementing multicast functionality at the logical layer. The objective of the OM problem is to construct an OM tree that minimizes transmission costs, including maximizing bandwidth and minimizing total delay and packet loss rate. The problem of constructing an OM tree with minimal transmission cost is equivalent to the minimum Steiner tree problem in graph theory. This problem aims to connect specified terminal nodes in a graph while minimizing transmission costs [41].

In this paper, we model the underlying network topology in SDN as an undirected connected graph $\mathfrak{G}=(\mathcal{V},\mathcal{E})$, where $\mathcal{V}=\{v_i\}$ denotes the finite set of nodes in the topology, $v_i$ represents the network nodes of the underlying topology, and $n_v=|\mathcal{V}|$ indicates the number of network nodes.

$\mathcal{E}=\{\langle v_i,v_j\rangle | v_i,v_j\in\mathcal{V},\ i\neq j\}$ signifies the set of links present in the network topology, with $\langle v_i,v_j\rangle$ representing the edge between node $v_i$ and node $v_j$. $n_e=|\mathcal{E}|$ denotes the number of edges in the network topology. Since we deal with an undirected graph, $\langle v_i,v_j\rangle$ and $\langle v_j,v_i\rangle$ represent the same edges. We assume that the network topology remains unchanged during multicast, meaning the number of nodes and the connections between edges do not alter during the multicast process.

Let $\mathcal{V}^s=\{v_{s_i}\}\subseteq\mathcal{V}$ denote the set of source nodes for overlay multicast, where $v_{s_i}\in\mathcal{V}$ represents the source node, the originating node initiating multicast data transmission. In this scenario, the number of source nodes is denoted as $n_{vs}=|\mathcal{V}^s|=1$. For simplicity, the source node is directly represented as $v_s$. Let $\mathcal{V}^d=\{v_{d_i}\}\subseteq\mathcal{V}$ denote the set of destination nodes for OM, where $v_{d_i}\in\mathcal{V}$ represents the destination node, the terminal node receiving data transmission initiated by the source node. $n_{vd}=|\mathcal{V}^d|$ denotes the number of destination nodes.

At the application layer, the set of OM source nodes, $\mathcal{V}^s$, and the set of destination nodes, $\mathcal{V}^d$, are abstracted into a logical topology, the overlay network, represented as $\mathfrak{G}^r=(\mathcal{V}^r,\mathcal{E}^r)$. Here, the overlay network nodes are $\mathcal{V}^r=\{v_i^r\}=\mathcal{V}^s\cup\mathcal{V}^d\subseteq\mathcal{V}$, and the edges of the overlay network are $\mathcal{E}^r=\{\langle v_i^r,v_j^r\rangle | v_i^r,v_j^r\in\mathcal{V}^r, i\neq j\}$. Specifically, $\langle v_i^r,v_j^r\rangle$ denotes a path from source node $v_i^r$ to destination node $v_j^r$. Within the overlay network, this is viewed as a virtual link from $v_i^r$ to $v_j^r$, termed an overlay link. This structure ensures that $\mathfrak{G}^r$ forms an undirected complete graph. $\mathfrak{G}^r$ The complete graph property ensures that any pair of nodes in the overlay network can communicate directly. The overlay link $\langle v_i^r,v_j^r\rangle$ corresponds to a unicast routing path $P(\langle v_i^r,v_j^r\rangle)=\langle v_i^r,v_{r_1},v_{r_2},\ldots,v_{r_k},v_j^r\rangle$ in the underlying network $\mathfrak{G}$, where $v_i\in\mathcal{V}$. To facilitate subsequent formula derivation, the path $P(\langle v_i^r,v_j^r\rangle)$ is transformed from a node sequence into an edge set as shown in Eq (1) below:

$$\hat{P}(\langle v_i^r,v_j^r\rangle)\triangleq\{\langle v_{r_l},v_{r_{l+1}}\rangle | l=0,1,\ldots,k\} \tag{1}$$

where $v_{r_0}=v_i^r, v_{r_{k+1}}=v_j^r$, which will be abbreviated as $\hat{P}_{ij}$.

The OM tree $\mathcal{T}$ is a spanning tree rooted at node $v_s$ on the overlay layer $\mathfrak{G}^r$, where $\mathcal{T}=(V^T,E^T)$, $V^T=V^r, E^T\subseteq\mathcal{E}^r$. Its overlay cost is defined as the sum of the costs of all underlying unicast routing paths corresponding to the links in the overlay layer, as shown in Eq (2):

$$C(\mathcal{T})=\sum_{\langle v_i^r,v_j^r\rangle\in\mathcal{T}} f(\langle v_i^r,v_j^r\rangle) \tag{2}$$

where $f(\langle v_i^r,v_j^r\rangle)$ represents the edge weight of the overlay network, abbreviated as $f_{ij}$, which is the transmission cost corresponding to the underlying path $\hat{P}(\langle v_i^r,v_j^r\rangle)$. Its value is determined by the path's remaining bandwidth, path latency, and path packet loss rate, as explained in detail below.

$bw_{ij}$ denotes the remaining bandwidth of the path $\hat{P}(\langle v_i^r,v_j^r\rangle)$, i.e., the minimum remaining bandwidth along the path from node $v_i^r$ to node $v_j^r$, is defined as shown in Eq (3):

$$bw_{ij} = \min_{\langle v_{r_l}, v_{r_{l+1}} \rangle \in \hat{P}_{ij}} \left( bw\left( \left\langle v_{r_l}, v_{r_{l+1}} \right\rangle \right) \right) \tag{3}$$

where $bw\left(\left\langle v_{r_l}, v_{r_{l+1}} \right\rangle\right)$ represents the remaining bandwidth between nodes $v_{r_l}$ and $v_{r_{l+1}}$ in $\mathfrak{G}$.

$delay_{ij}$ is the total delay of path $\hat{P}\left(\left\langle v_i^r, v_j^r \right\rangle\right)$, representing the sum of all link delays in the path, defined as shown in Eq (4):

$$delay_{ij} = \sum_{\langle v_{r_l}, v_{r_{l+1}} \rangle \in \hat{P}_{ij}} delay\left( \left\langle v_{r_l}, v_{r_{l+1}} \right\rangle \right) \tag{4}$$

where $delay\left(\left\langle v_{r_l}, v_{r_{l+1}} \right\rangle\right)$ is the latency between nodes $v_{r_l}$ and $v_{r_{l+1}}$ in $\mathfrak{G}$.

$loss_{ij}$ is the packet loss rate of path $\hat{P}\left(\left\langle v_i^r, v_j^r \right\rangle\right)$, defined as shown in Eq (5) below:

$$loss_{ij} = 1 - \prod_{\langle v_{r_l}, v_{r_{l+1}} \rangle \in \hat{P}_{ij}} \left( 1 - loss\left( \left\langle v_{r_l}, v_{r_{l+1}} \right\rangle \right) \right) \tag{5}$$

The optimization objective of the minimum cost $f_{ij}$ is to maximize the path's remaining bandwidth $bw_{ij}$ while minimizing the path's total delay $delay_{ij}$ and packet loss rate $loss_{ij}$, as shown in Eq (6):

$$f_{ij} = \beta_1 \left( 1 - bw_{ij} \right) + \beta_2 delay_{ij} + \beta_3 loss_{ij} \tag{6}$$

where $\beta_k, k = 1,2,3$ denote the weighting coefficients of path bandwidth, delay, and packet loss rate, respectively, subject to the constraint: $\sum_{i=1}^{3} \beta_i = 1$. The core objective of weight design is that multi-QoS metric optimization outperforms single-QoS metric optimization. Considering the transmission characteristics of low delay and high reliability ensures end-to-end QoS. By adjusting the proportions of different weights, congested paths can be effectively avoided, so that low-delay and high-reliability links are selected for transmission, thereby adapting to and satisfying the requirements of vaious application scenarios and improving the overall path service quality. The values of the weighting coefficients obtained through repeated adjustments and tests in this study are presented in the subsequent experimental section.

The optimal overlay multicast tree involves finding a minimum spanning tree that minimizes the sum of overlay costs across all spanning edges. The model for the optimal overlay multicast tree problem is shown in Eq (7):

$$\min\ C(\mathcal{T}) = \sum_{\langle v_i^r, v_j^r \rangle \in \mathcal{T}} f\left( \left\langle v_i^r, v_j^r \right\rangle \right) \tag{7}$$

*3.2. Problem decomposition*

Constructing an overlay multicast tree is essentially a Steiner tree problem with QoS constraints. To circumvent the NP-completeness of the minimum Steiner tree problem, we propose a sequence-driven, two-stage heuristic problem framework: The first stage involves sequence generation, producing the destination node sequence $S$ and dynamically narrowing the selection range of parent nodes; the second stage involves tree expansion, iteratively extending the multicast tree $\mathfrak{T}$ based on sequence order. This sequence-driven approach controls topology coherence, transforming the global optimization problem into a locally cost-minimizing decision chain.

Let $\mathcal{P}\left(\mathcal{V}^d\right)$ denote the set of all permutations of the destination node set $\mathcal{V}^d$. An element $S \in \mathcal{P}\left(\mathcal{V}^d\right)$ in $\mathcal{P}\left(\mathcal{V}^d\right)$ is represented as $S=(n_1, n_2, n_3, \ldots, n_{n_{vd}}), n_i \in \mathcal{V}^d, \forall i \in \{1,2,3,\ldots,n_{vd}\}$. The sequence $S$ transforms combinatorial optimization into a sequential decision chain by controlling the candidate set of preferred data forwarding source nodes $\mathcal{S}_k = \mathcal{V}^s \cup \{n_1, n_2, \ldots, n_{k-1}\}$, thereby simplifying the decision space. When expanding the tree $\mathfrak{T}$, a local optimization problem $f(P_s(\mathcal{S}_k, n_k))$ is solved for each $n_k$. Here, $P_s(\mathcal{S}_k, n_k)$ denotes a unicast routing path constructed in the underlying network $\mathfrak{G}$, with a source node selected from the source node candidate set $\mathcal{S}_k$ and a destination node $n_k$. The mapping $\mathcal{O}(\mathcal{X}) = x, x \in \mathcal{X}$ is defined to select an element from the set. In this paper, $\mathcal{O}\left(\mathcal{S}_k\right)$ represents the selected destination node, so $P_s(\mathcal{S}_k, n_k)$ can also be expressed as $P\left(\left\langle \mathcal{O}\left(\mathcal{S}_k\right), n_k \right\rangle\right)$.

In summary, the OM tree optimization model can be further extended, expressed as shown in Eq (8):

$$\underset{S \in \mathcal{P}\left(\mathcal{V}^d\right)}{\arg\min} \sum_{n_k \in S, \mathcal{S}_k \subseteq \mathcal{V}^d} f\left(P\left(\left\langle \mathcal{O}\left(\mathcal{S}_k\right), n_k \right\rangle\right)\right) \tag{8}$$

Therefore, the overlay multicast construction process in this paper consists of two-tier optimization tasks:

1) High-level policy: Responsible for sequence generation objective learning. Within the set of destination nodes $\mathcal{V}^d$, generate a sequence $S=(n_1, n_2, n_3, \ldots, n_{n_{vd}}), n_i \in \mathcal{V}^d, \forall i \in \{1,2,3,\ldots,n_{vd}\}$ with the optimization goal of minimizing the OM tree cost $C\left(\mathcal{T}\right)$.

2) Low-level strategy: Iteratively finds the optimal overlay network parent node for each destination node $v_{d_i}$ from the sequence and constructs the corresponding optimal routing path. The optimization objective for each iterative subtask is to minimize the routing path cost function $f\left(P\left(\left\langle \mathcal{O}\left(\mathcal{S}_k\right), n_k \right\rangle\right)\right)$.

## 4. Model architecture

The algorithmic architecture for SDN-based multi-agent OM Routing designed in this paper is illustrated in Figure 1.

### *4.1. Data plane*

The data plane is a critical component of the SDN architecture, responsible for processing and forwarding packets within the SDN network according to routing policies issued by the control plane. It is also referred to as the forwarding plane. This plane consists of a series of physical devices with strong forwarding capabilities and simplified control logic, such as programmable switches and routers. Within the SDN architecture, the data plane lacks autonomous control capabilities. Instead, it relies entirely on the upper-layer controller to issue unified scheduling through flow table rules via the southbound OpenFlow protocol interface, thereby achieving a flexible and controllable data flow forwarding mechanism. Furthermore, the data plane reports current network state information to the control plane via the Southbound Interface based on controller instructions. This includes port traffic statistics (e.g., received and transmitted packet counts and byte counts) and packet drop events, providing foundational support for efficient packet forwarding by supplying comprehensive network link status data.

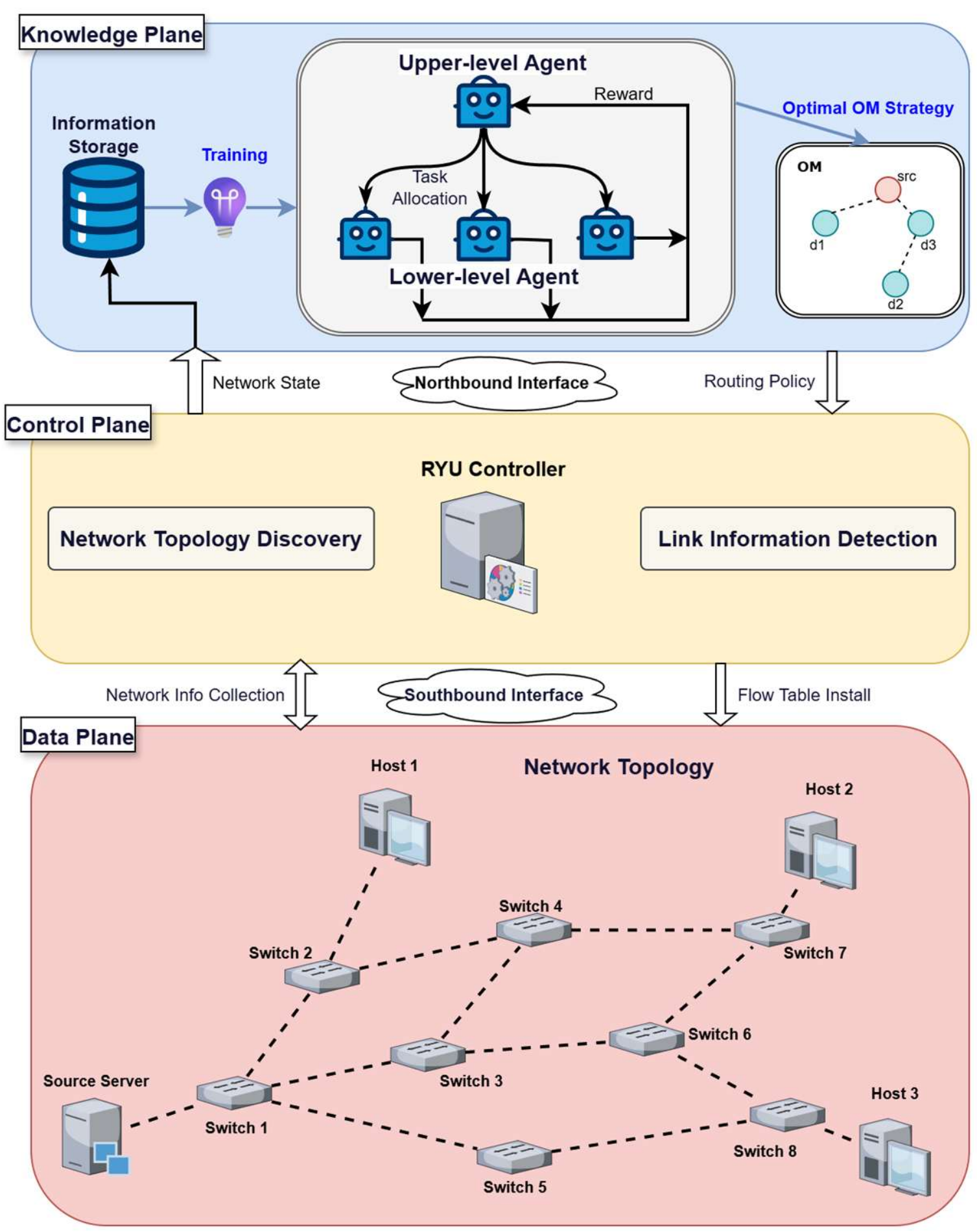

**Figure 1.** Model architecture.

*4.2. Control plane*

The control plane occupies a central position within the SDN architecture, responsible for managing the global network state, defining forwarding policies, and dynamically scheduling resources. The controller interacts with data plane switches via the southbound OpenFlow protocol to discover network topology, collect network state information, and construct a global network view. During initial connection, the controller sends an OFPT_FEATURES_REQUEST message to obtain the switch's capability information and receives the switch's OFPT_FEATURES_REPLY response to acquire details such as the switch's Datapath ID, protocol version, port list, and attributes. After acquiring switch information, the control plane further discovers the network topology using the Link Layer Discovery Protocol (LLDP). The controller actively sends LLDP probe packets carrying sender identification to switch ports via Packet-Out messages. Upon receiving these packets, switches flood the LLDP packets from non-controller ports to adjacent devices. When adjacent devices receive LLDP packets, they trigger Packet-In messages due to no matching flow table entries. These messages report the LLDP packet and its ingress port information to the controller. The controller parses the LLDP source information within the Packet-In message against the reported switch ID and ingress port, thereby establishing the connection relationship “Switch A - Port X → Switch B - Port Y”. This process progressively integrates to form a global topology view.

During the controller’s collection of network state information, it periodically sends statistical request messages (OFPT_STATS_REQUEST) to each switch. The switches respond with their port operational status and flow table statistics (OFPT_PORT_STATS_REPLY). The raw data returned from each port typically includes the number of transmitted packets ( $txp$ ), the number of received packets ( $rxp$ ), the number of transmitted bytes ( $txb$ ), the number of received bytes ( $rxb$ ), and the duration over which bytes are transmitted ( $tdur$ ). Based on this raw port data, the controller can further compute link performance metrics.

The residual bandwidth $bw(\langle v_i, v_j \rangle)$ of link $\langle v_i, v_j \rangle$ is defined as the difference between its maximum bandwidth $bw_{\max}$ and its utilized bandwidth $bw_{used}$. The controller computes use by dividing the difference in transmitted bytes between two consecutive port statistics messages by the corresponding time interval. Specifically, $bw_{used}$ can be derived from $txb$, $rxb$, and $tdur$ at both ends of the link, as shown in Eq (9).

$$\begin{aligned} bw(\langle v_i, v_j \rangle) &= bw_{\max} - bw_{used} \\ &= bw_{\max} - \frac{\left| (txb_i + rxb_i) - (txb_j + rxb_j) \right|}{tdur_j - tdur_i} \end{aligned} \tag{9}$$

where ` $txb_*$ ` and ` $rxb_*$ ` represent the number of bytes transmitted, and ` $tdur_*$ ` represents the duration of transmission for Node* in the network.

The link packet loss rate $loss(\langle v_i, v_j \rangle)$ can be calculated based on the number of transmitted and discarded packets per port. Taking the transmission direction as an example, the packet loss rate formula is shown in Eq (10). Here, $txp_*$ and $rxp_*$ represent the number of transmitted packets from Node* in the network

$$loss(\langle v_i, v_j \rangle) = \frac{txp_i - rxp_j}{txp_i} \tag{10}$$

In SDN, packet communication between switches requires controller-mediated forwarding. Therefore, link latency $delay(\langle v_i, v_j \rangle)$ must be approximated via the LLDP protocol. The unidirectional link latency is calculated from the total transmission delay of LLDP packets $\mathrm{T_{lldp}}$ and the round-trip delay (RTT) between the controller and the switch, as shown in Eq (11):

$$delay(\langle v_i, v_j \rangle) = \frac{T_{lldpi} + T_{lldpj} - RTT_i + RTT_j}{2} \tag{11}$$

Here, $T_{lldp*}$ represents the LLDP latency in the direction originating from node*, while $RTT_*$

denotes the round-trip latency from the controller to node*.

The control plane constructs a unified global network view using the network data derived from these calculations. This provides an abstract representation of network resources for upper-layer applications, enabling the knowledge plane, which interfaces with the application layer, to formulate policies based on overall network state for global optimization.

### *4.3. Application plane*

The application plane resides at the topmost layer of the SDN architecture, and is primarily responsible for implementing specific network functions and service policies. It collaborates with the knowledge plane to enhance network decision-making capabilities and efficiency. By transmitting service requirements and network state information to the knowledge plane, and leveraging the global network view and programmable abstract interfaces provided by the control plane, the application plane distributes executable optimization policies, fed back by the knowledge plane, to the data plane for execution via the controller.

### *4.4. Knowledge plane*

The knowledge plane serves as an extension layer for intelligent decision-making in the new SDN architecture. Positioned above the control plane and operating in tandem with the application plane, its core function is to autonomously optimize network policies through intelligent algorithms. The multi-agent reinforcement learning overlay multicast algorithm described in this paper operates within the knowledge plane.

The upper-layer agent in the hierarchical reinforcement learning algorithm for the knowledge plane dynamically decomposes the overlay multicast coverage objectives based on application-layer requirements. Utilizing global network topology information, it splits these objectives into multiple local sub-goals and assigns them to lower-layer agents. Multiple agents perform parallel computations based on their respective sub-goals using the multidimensional traffic matrix derived from network state information (including remaining bandwidth, link latency, and packet loss rate) collected by the control plane. They collaboratively construct the overlay multicast tree and then issue instructions to the control plane, which in turn disseminates flow table entries to the data plane.

## 5. MA-DHRL-OM algorithm design

Building upon the sequence-driven two-stage heuristic problem framework proposed in Section 3, we design a hierarchical deep reinforcement learning framework to address the problem.

### *5.1. MA-DHRL-OM algorithm design*

The proposed MA-DHRL-OM algorithm consists of two hierarchical layers. In the upper layer, the Proximal Policy Optimization (PPO) algorithm is employed to optimize the first-stage decision variable: The destination node sequence $S$. This sequence $S$ is then partitioned into $n_{vd}$ sub-goals through goal decomposition. In the lower layer, a single agent $agent_i$ governed by the Soft Actor-Critic (SAC) algorithm is assigned to each destination node $v_{d_i}$ to perform parent node selection and construct intelligent routing paths. The multi-agent system in the lower layer operates collaboratively and in parallel to build the overlay multicast tree.

### *5.2. Upper-layer RL algorithm: PPO*

#### 5.2.1. State space design

The state space of reinforcement learning agents largely determines their environmental perception and understanding capabilities. Upper- and lower-layer reinforcement learning agents in this paper employ multi-channel matrix state spaces, enabling comprehensive environmental awareness. Through integrated convolutional layers, agents can recognize feature representations across dimensions.

The state space of the upper-layer deep reinforcement learning agent PPO is a three-channel matrix: $X_{PPO}$. This matrix is composed of three stacked layers: The Node Distance Matrix (NDM), the Source-to-Node Distance Matrix (SNDM), and the Node Selection Matrix (NSM), as illustrated in Figure 2.

The node distance matrix $NDM = \left[ ndm_{ij}^{norm} \right] \in R^{n_{vd} \times n_{vd}}$ represents the distance adjacency matrix for the overlay layer. Here, $ndm_{ij}^{norm}$ denotes the normalized minimum hop count distance between the OM destination nodes $v_i^r$ and $v_j^r$ in the current underlying network topology.

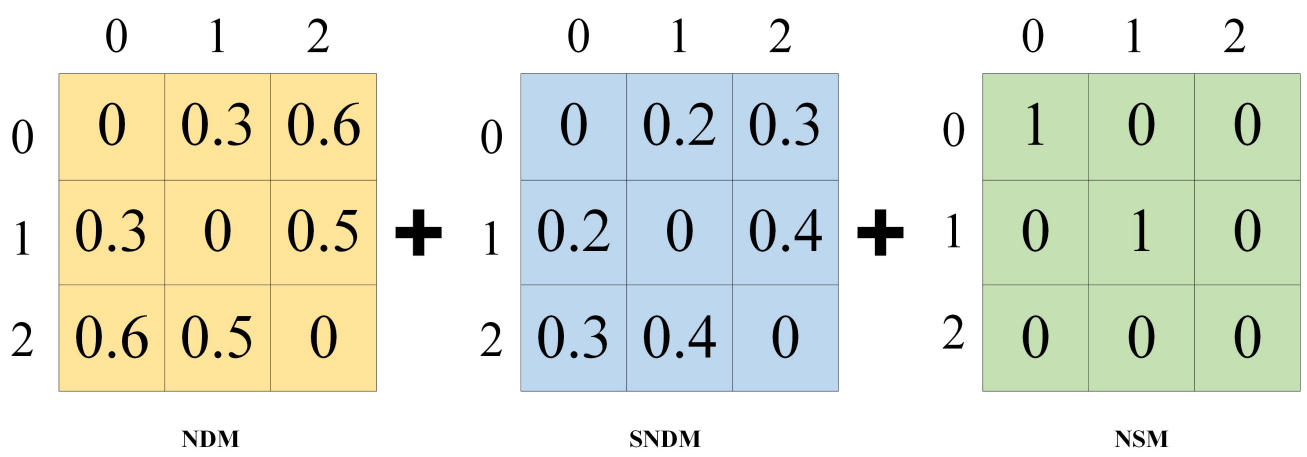

**Figure 2.** State space representation of the upper-layer reinforcement learning algorithm.

The source-destination distance matrix $SNDM = \left[ sndm_{ii}^{norm} \right] \in R^{n_{vd} \times n_{vd}}$ is a diagonal matrix, where the values $sndm_{ii}^{norm}$ on the diagonal represent the normalized shortest hop count distance from the OM source node $v_s$ to each destination node $v_{d_i}$ in the current underlying network topology.

The node selection matrix $NSM = \left[ nsm_{ij}^{norm} \right] \in R^{n_{vd} \times n_{vd}}$ is a diagonal matrix. The values on the diagonal $nsm_{ii}^{norm}$ indicate whether $v_{d_i}$ has been selected. If not yet selected, $nsm_{ii}^{norm} = 1$; otherwise, $nsm_{ii}^{norm} = 0$.

#### 5.2.2. Action space design

To enable the upper-layer agent to sort all destination nodes and output a sequence $C$, we design the action space $\mathcal{A}^{\text{PPO}} = \{ a_i^{ppo} \}, \left| \mathcal{A}^{\text{PPO}} \right| = n_{vd}$ for the upper-layer PPO algorithm. $n_{vd}$ represents the number of destination nodes, $a_i^{ppo}$ denotes the currently selected destination node $v_{d_i}$, and $a_i^{ppo}$ indicates the destination node $v_{d_i}$ selected in the current state.

To accelerate the upper-layer agent's efficiency and enable it to complete the goal node sorting within time steps without repetition, we design a one-dimensional dynamic mask vector $\mathrm{Mask}^{PPO}$. When a goal node $v_{d_i}$ has been selected, the corresponding index position $i$ in the dynamic mask vector $\mathrm{Mask}^{PPO}$ is set to false, while all other elements remain true (at the start of a round, $\mathrm{Mask}^{PPO}$ is a one-dimensional Boolean vector with all elements set to true). When the policy network of the PPO algorithm outputs the raw prediction values $logist$ of the discrete action probability distribution under the current state, these values are multiplied by the corresponding $\mathrm{Mask}^{PPO}$ before undergoing SoftMax and probability distribution sampling. This technique of local observation enables the PPO agent to avoid selecting already-ranked nodes during decision-making, ensuring the output of a decision sequence $C$ for the destination node within a single round.

### 5.2.3. Reward function design

Rewards serve as feedback signals received by the agent from the environment, measuring the quality of state transitions resulting from executed actions. They constitute the core mechanism driving the agent's learning of optimal policies, with their design directly impacting learning efficiency and the quality of the final policy.

In the upper-layer PPO algorithm, the reward function comprises two components: intermediate reward and sequence reward.

The intermediate reward $R_{Intm}$ is granted by the upper-layer agent PPO after selecting a destination node. At the start of a round, when initially selecting a destination node $v_{r_i}$, the intermediate reward $R_{Intm} = -sndm_{ii}^{norm}$ is the normalized shortest hop count distance from the source to the destination node $v_{r_i}$. When not at the start of a round, during node selection from the destination node $v_{r_i}$ to $v_{r_j}$, $R_{Intm} = -ndm_{ij}^{norm}$. We employ negative node distance consumption as the intermediate reward for the upper-layer destination node ranking algorithm, guiding agents to prioritize topologically relevant connections.

The sequence reward $R_{Seq}$ is the round-end reward for the upper-layer agent PPO after completing the target phase ranking. The sequence $C$ of the agent's final ranking result will be decomposed into $n_{vd}$ sub-tasks. These are assigned to $n_{vd}$ SAC RL at the lower layer, each corresponding to a sub-goal. The lower-layer multi-agent system will first complete the parent node selection for sub-goals and then the intelligent routing path construction. The results are then computed and returned as rewards to the upper layer, as shown in Eq (12).

$$R_{Seq} = \sum_{v_{r_i} \in v_d} R_{FB_i} \tag{12}$$

$R_{FB_i}$ is the reward that the agent corresponding to destination node $v_{r_i}$ feeds back to the upper-level agent in the current episode, while $R_{Seq}$ represents the cumulative sum of feedback rewards from all agents. The feedback reward $R_{FB_i}$ for $agent_i$ is computed as shown in Eq (13).

$$R_{FB_i} = \sum_{link_k \in path_i} R_{\mathrm{link}_k} \tag{13}$$

Here, $link_k$ denotes a link in the routing path $path_i = \{link_k\}, k = 1,2,\ldots, length_i$ constructed by $agent_i$, and $length_i$ represents the length of $path_i$. The $R_{\mathrm{link}_k}$ is the composite cost-based reward for $link_k$, which will be formally defined in the reward function design of the lower-layer SAC algorithm; thus, it is not elaborated here.

5.2.4. Parameter updates

The PPO algorithm updates the parameters of the policy network and value network by optimizing a clipped objective function. This objective function typically consists of two parts: The advantage function $A_t$ and the probability ratio $r(\theta)$. $r(\theta)$ represents the ratio of the probability of selecting an action $a_t$ under the current policy $\pi_\theta$ to the probability under the old policy $\pi_{\theta_{old}}$, as shown in Eq (14).

$$r(\theta) = \frac{\pi_\theta(a_t \mid s_t)}{\pi_{\theta_{old}}(a_t \mid s_t)} \tag{14}$$

Here, $\pi_\theta(a_t \mid s_t)$ denotes the probability of the current policy selecting action $a_t$ under state $s_t$, and $\pi_{\theta_{old}}(a_t \mid s_t)$ is the probability of the old policy selecting the same action under the same state. The ratio $r(\theta)$ measures the magnitude of the policy update. Building upon this, the PPO algorithm employs a clipping mechanism to restrict the extent of policy updates, and its objective function is formulated as shown in Eq (15).

$$\mathcal{L}^{CLIP}(\theta) = \mathbb{E}_t\left[\min\left(r(\theta)A_t, clip\left(r(\theta), 1-\varepsilon, 1+\varepsilon\right)A_t\right)\right] \tag{15}$$

Among these, $A_t$ represents the estimated value of the advantage function for the state-action pair $(s_t, a_t)$, computed using Generalized Advantage Estimation (GAE). The function $clip(x,l,r) = \max\left(\min(x,r), l\right)$ constrains $x$ within the range $[l,r]$, where $\varepsilon$ is a hyperparameter of $clip$ representing the clipping range, typically set to 0.1 or 0.2. The design of $clip$ and $\min$ ensures that the objective function avoids overly large step updates during a policy update, achieving stable policy improvement and preventing policy oscillations that may occur in traditional policy gradient methods.

The parameters of the value network ($\phi$) are typically updated independently, aiming to minimize the error between the predicted state value and the target value. The mean squared error (MSE) is used to optimize the value function parameters, as shown in Eq (16):

$$\mathcal{L}^V(\phi) = \mathbb{E}_t\left[\left(V_\phi(s_t) - V^{\mathrm{target}}(s_t)\right)^2\right] \tag{16}$$

where $V_\phi(s_t)$ represents the state value function predicted by the value network, and $V^{\mathrm{target}}(s_t)$ denotes the target value function.

*5.3. The lower-layer multi-agent RL algorithm: SAC*

5.3.1. State space design

The state space for the lower-layer SAC Algorithm is a multi-channel matrix, $X_{SAC}$, composed of a three-channel traffic matrix (TM) and a node location matrix (LOC) stacked together, as shown in Figure 3.

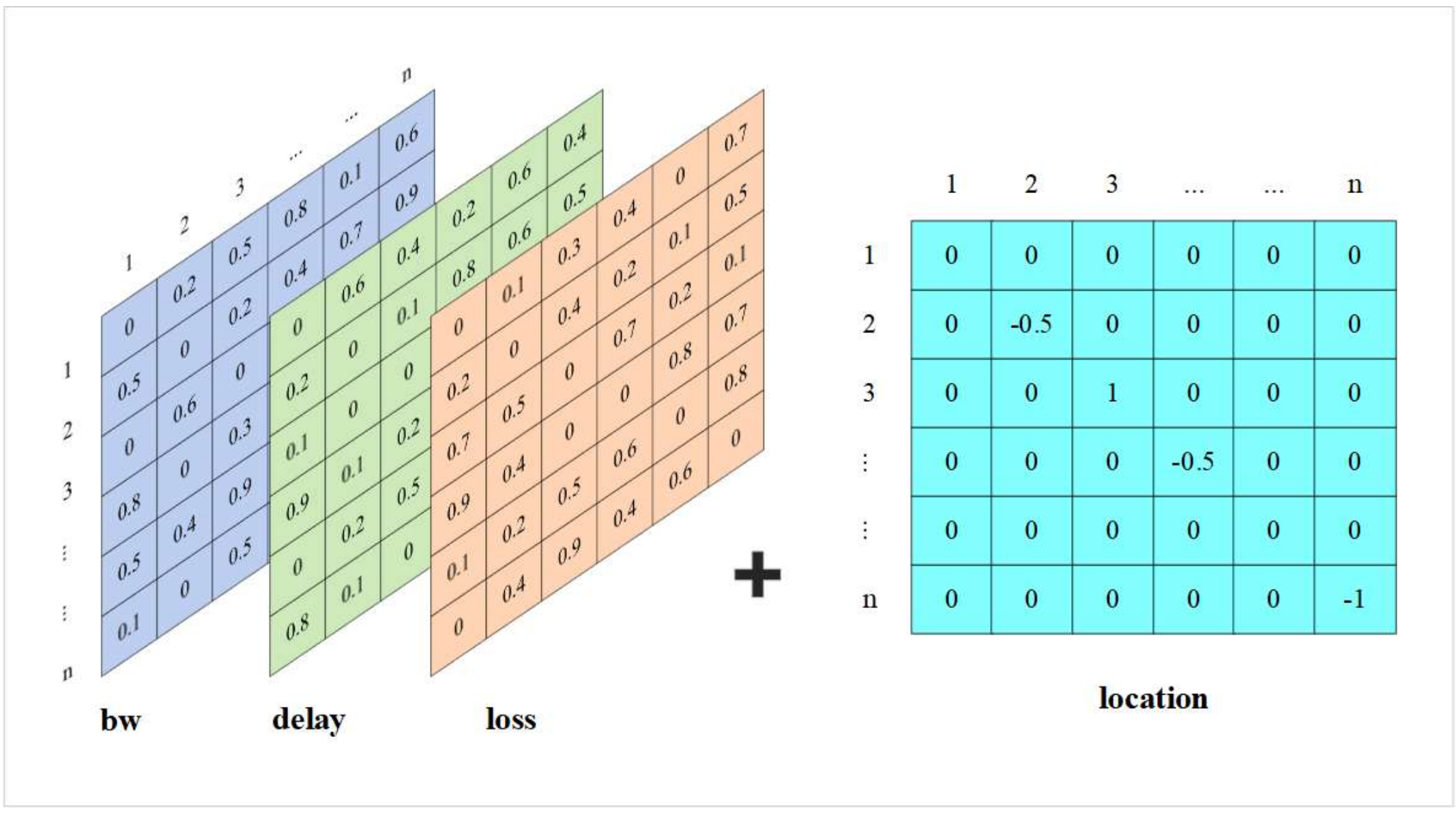

**Figure 3.** State space representation of the lower-layer reinforcement learning algorithm.

TM is constructed by normalizing network state information collected from the data plane. It comprises the normalized residual bandwidth matrix ($M_{bw}^{norm} = \left[ bw_{ij}^{norm} \right] \in R^{n_v \times n_v}$), the normalized link latency matrix ($M_{delay}^{norm} = \left[ delay_{ij}^{norm} \right] \in R^{n_v \times n_v}$), and the normalized packet loss rate matrix ($M_{loss}^{norm} = \left[ loss_{ij}^{norm} \right] \in R^{n_v \times n_v}$). All three matrices are adjacency symmetric matrices of size $n_v \times n_v$, where $n_v$ is the number of nodes in the network topology $\mathfrak{G}$. Each element in the matrix represents the normalized network state between the nodes corresponding to its respective row and column coordinates. For example, in the matrix $M_{bw}^{norm}$, the value $bw_{ij}^{norm}$ corresponds to the normalized remaining bandwidth between network nodes $v_i$ and $v_j$ in the underlying topology $\mathfrak{G}$. If the two nodes are not connected, then $bw_{ij}^{norm} = 0$.

LOC is defined as $M_{loc} = \left[ loc_{ij} \right] \in R^{n_v \times n_v}$, marking the positions of the current routing node, the candidate departure node for the target node under the current task, and the target node. The values of $location_{ij}$ are calculated as shown in Eq (17).

$$location_{ij} = \begin{cases} 1, i = j = loc_{\text{now}} \\ -1, i = j = loc_{\text{target}} \\ -0.5, i = j = loc_{\text{cand}} \\ 0, else \end{cases} \tag{17}$$

where $loc_{\text{now}}$ denotes the node position index of the current agent, $loc_{\text{target}}$ denotes the destination node index of the agent, and $loc_{\text{cand}}$ denotes the position index of the candidate node. When $loc_{\text{now}} = loc_{\text{target}}$, this indicates that the agent has successfully reached the destination node position, signifying algorithm termination.

### 5.3.2. Action space design

To enable the agent to simultaneously perform node selection and unicast routing path construction, the action space for the underlying SAC algorithm is designed as follows: $\mathcal{A}^{SAC} = \{a_i^{sac}\}, |\mathcal{A}^{SAC}| = n_{\max}$. $n_{\max}$ represents the maximum degree generated by interconnecting the source node $v_s$ and all destination nodes $v_{d_i}$ in the graph $\mathfrak{G}$. $a_i^{sac}$ denotes the traversal action taken by the SAC agent at the node corresponding to its current state $s_t$. Actions differ across nodes, with the next node position determined by the neighbor list.

To prevent invalid actions from affecting the agent, we introduce a mask mechanism in the action selection component of the SAC agent algorithm. After the source node $v_s$ and all destination nodes $v_{d_i}$ are interconnected, and when the agent is at a node $v_i$ that is not the starting node, the actual neighboring nodes connected to node $v_i$ via links are marked as True in the mask vector. Invalid actions in the action space that exceed the actual neighbor range of this node are marked as False in the mask vector.

When the agent is at a node $v_i$, which is the departure node, the action space for this node's state includes not only the mapping of original neighbors but also the selection of destination nodes. This enables the agent to choose different departure nodes to construct routing paths. Its mask vector sets true values not only at the corresponding positions of connected neighbor nodes as described above but also at the positions of candidate destination nodes under the task constraints of the current round. The remaining positions of the mask are set to false. This prevents the generation of invalid actions.

### 5.3.3. Reward function design

In the lower-layer SAC reinforcement learning algorithm, the rewards designed in this paper are categorized into immediate rewards and terminal rewards.

The immediate reward is numerical feedback returned by the environment to the agent immediately after the agent executes an action. When the agent is at a candidate node, and the selected action points to another candidate node, this action constitutes a parent node switch within the overlay layer. The instantaneous reward at this point is defined as $R_{imm1}$, as shown in Eq (18):

$$R_{imm1} = R_{step} \tag{18}$$

where $R_{step} = -0.1$ represents the single-step consumption value of the SAC agent. The instantaneous reward used for the 'node switch' action is a low-cost consumption reward. It guides the agent to switch starting nodes at near-zero cost, intelligently selecting the optimal parent node for the destination node. The setting of $R_{step}$ ensures that $R_{imm1}$ is not entirely zero-cost. $R_{step}$ acts as a "time cost" penalty, preventing the agent from adopting conservative or inefficient strategies due to zero cost, thereby avoiding local deadlocks and ensuring long-term optimization of the routing path.

When an agent's action is effective and not a "node switch" action, the instantaneous reward is $R_{imm2}$, as shown in Eq (19):

$$R_{imm2} = R_{link} + R_{step} \tag{19}$$

Here, $R_{link}$ represents the link composite cost, previously defined and used in the reinforcement learning PPO algorithm reward function design. Its calculation is shown in Eq (20):

$$R_{link} = \beta_1(bw_{ij}^{norm} - 1) - \beta_2 delay_{ij}^{norm} - \beta_3 loss_{ij}^{norm} \tag{20}$$

where $bw_{ij}^{norm}$ , $delay_{ij}^{norm}$ , and $loss_{ij}^{norm}$ represent the normalized residual bandwidth, link latency, and packet loss rate values, respectively, of the link $\langle v_i, v_j \rangle$ added to the routing path chain following the state transition $s_t \to s_{t+1}$ after the SAC agent performs the action $a_t$. These values guide the agent to seek actions within the current state that point toward links with greater residual bandwidth and lower delay and packet loss rates. As evident from the above formula, the instantaneous rewards $R_{imm1}$ and $R_{imm2}$ are always negative. The agent thus prioritizes reaching the target as quickly as possible to minimize cumulative loss, enabling the underlying reinforcement learning algorithm to output a routing path with minimal total link cost.

The termination reward $\mathrm{R_{Trm}}$ is the reward value received when the agent reaches a specific target state or failure state during exploration. When the agent reaches the destination node within the specified time step $failsteps$, a positive reward is given, such as $\mathrm{R_{Trm}} = 10$. An appropriate positive value of $\mathrm{R_{Trm}}$ can enable the agent to learn the final objective of the task and guide it to reach the goal more quickly. When the agent fails to successfully reach the target node position after exceeding fail steps time steps, a relatively large negative penalty is imposed; for example, $\mathrm{R_{Trm}} = -10$. $failsteps$ represents the failure time steps, and its calculation formula is $failsteps = 1.5 n_{\mathrm{v}}$. The setting of a 1.5-fold ratio introduces a time cost constraint, preventing the agent from attempting random actions indefinitely, thereby forcing efficient learning of the agent under sufficient exploration space.

### 5.3.4. Parameter updates

SAC is an offline policy algorithm based on maximum entropy reinforcement learning. It optimizes the policy to maximize the weighted sum of expected cumulative reward and policy entropy. It employs five neural network architectures for function approximation: One parameterized policy network (Actor), two state-action value function networks (Critics), and two corresponding target Q-networks. The actor network $\pi_\phi$ outputs a probability distribution defined over the finite action set $\mathcal{A}$ , denoted as $\pi_\phi(a \mid s) \in [0,1]^{|\mathcal{A}|}$ , where $|\mathcal{A}|$ represents the action space dimension. Two parallel critics networks $(Q_{\theta_1}, Q_{\theta_2})$ output the Q-value vector $Q_\theta(s) \in \mathbb{R}^{|\mathcal{A}|}$ for all possible actions under state $s$ .

The parameter update process centers on maximizing the cumulative reward objective with entropy regularization, defined by the objective function shown in Eq (21).

$$\mathcal{J}(\pi) = \mathbb{E}_{(s_t, a_t) \sim \rho_\pi}\left[\sum_{t=0}^{T} \gamma^t \left( r(s_t, a_t) + \alpha \mathcal{H}\left(\pi(\cdot \mid s_t)\right) \right)\right] \tag{21}$$

$\mathcal{H}(\pi) = -\mathbb{E}_{a \sim \pi}\left[log\pi(a \mid s)\right]$ denotes the policy entropy, $\alpha > 0$ represents the entropy regularization coefficient, and $\gamma \in [0,1]$ serves as the discount factor. This optimization objective is achieved through three network parameter updates.

Critic network parameter updates follow temporal difference learning principles, approximating the Q-function by minimizing bellman error. The target value is computed directly as the mathematical expectation of the policy distribution, avoiding variance introduced by discrete action sampling:

$$y_t = r_t + \gamma \mathbb{E}_{\alpha \sim \pi_\phi} \left[ \min_{j=1,2} Q_{\bar{\theta}_j} \left( s_{t+1}, a' \right) - \alpha \log \pi_\phi \left( a' \mid s_{t+1} \right) \right] \tag{22}$$

where $Q_{\bar{\theta}_j}$ denotes the target network output, and its parameter $\bar{\theta}_j$ is updated via Polyak averaging (soft update):

$$\bar{\theta}_j \leftarrow \tau \theta_j + \left( 1 - \tau \right) \bar{\theta}_j, \tau \ll 1 \tag{23}$$

The critic loss function is defined as the mean squared error between predicted and target values:

$$\mathcal{L}_Q \left( \theta_i \right) = \mathbb{E}_{\left( s_t, a_t, \tau_t, s_{t+1} \right) \sim \mathcal{D}} \left[ \frac{1}{2} \left( Q_{\theta_i} \left( s_t, a_t \right) - y_t \right)^2 \right], i = 1,2 \tag{24}$$

Here, $\mathcal{D}$ represents the experience replay pool, and $a_t$ denotes the actual executed action. Actor network parameter updates employ analytical calculation methods in discrete spaces. The policy network optimization objective is:

$$\mathcal{J}_\pi (\phi) = \mathbb{E}_{s_t \sim \mathcal{D}} \left[ \alpha \mathcal{H} \left( \pi_\phi \left( \cdot \mid s_t \right) \right) + \mathbb{E}_{a \sim \pi_\phi} \left[ \min_{j=1,2} Q_{\theta_j} \left( s_t, a \right) \right] \right] \tag{25}$$

where $\alpha \mathcal{H} \left( \pi_\phi \left( \cdot \mid s_t \right) \right)$ represents the entropy regularization term and $\mathbb{E}_{a \sim \pi_\phi} \left[ \min_{j=1,2} Q_{\theta_j} \left( s_t, a \right) \right]$ denotes the Q-value function expectation term. The objective function calculation in Eq (25) fully exploits the completeness axiom of discrete probability distributions $\sum_{a \in \mathcal{A}} \pi \left( a \mid s \right) = 1$ to achieve analytical derivation. The final policy loss function is defined as in Eq (26):

$$\mathcal{L}_\pi (\phi) = -\mathcal{J}_\pi (\phi) = \mathbb{E}_{s_t \sim \mathcal{D}} \left[ -\alpha \mathcal{H} \left( \pi_\phi \right) + \sum_a \pi_\phi \left( a \mid s_t \right) \min_j Q_j \left( s_t, a \right) \right] \tag{26}$$

The loss function with adaptive entropy coefficient update is given in Eq (27). When the policy entropy falls below the preset target $\mathcal{H}_{\text{target}}$, the training objective $\mathcal{L}(\alpha)$ increases the temperature coefficient $\alpha$. This enhances the weighting of the policy entropy term during the minimization of the policy loss function $\mathcal{L}(\alpha)$, encouraging increased exploration. Conversely, when the policy entropy exceeds $\mathcal{H}_{\text{target}}$, the training objective reduces $\alpha$, diminishing the entropy term's influence and directing the policy toward refining action value estimates during training.

$$\mathcal{L}(\alpha) = \mathbb{E}_{s_t \sim \mathcal{D}} \left[ -\alpha \left( \mathcal{H} + \mathcal{H}_{\text{target}} \right) \right] \tag{27}$$

*5.4. MA-DHRL-OM algorithm flow*

To simplify the pseudocode framework of the MA-DHRL-OM algorithm, we design a single-round training algorithm for a single SAC agent, as shown in Algorithm 1.

**Algorithm 1**: SAC agent training one episode

**Input:** Global network topology $\mathfrak{G}=(\mathcal{V},\mathcal{E})$, network state information matrix $\mathrm{TM}$, current set of sub-task starting nodes $srcs$, algorithm agent $agent_i$, batch update size $bs$

**Output:** Select an optimal node $s'$ from the departure node set $srcs$ and construct its optimal unicast routing path to the destination node $\mathrm{L}'$

1: reset environment state $s_t$ based on the srcs set and network information $\mathrm{TM}$, initialize path list $\mathrm{L}$
2: **While True do**
3: update $\mathrm{L}$ based on current node position
4: select action $a_t^{sac}=\pi_\phi(s_t)$ based on current policy of $agent_i$
5: get $mask_{s_t}$ from the $agent_i$ environment based on the state $s_t$
6: execute action $a_t^{sac}$, obtain reward $r_t$, state becomes $s_{t+1}$
7: place $(s_t, a_t^{sac}, r_t, s_{t+1}, done, mask_{s_t})$ into the experience replay pool $\mathcal{D}$
8: **If** the capacity of $\mathcal{D}$ $\geq$ the minimum training capacity **do**
9: Sampling $bs$ data tuples from $\mathcal{D}$ $\left\{\left(s_t^{(i)}, a_t^{(i)}, r_t^{(i)}, s_{t+1}^{(i)}, done, mask_t^{(i)}\right)\right\}_{i=1}^{bs}$
10: calculate the target value for the sample data tuple using Eq (22)
11: update the parameters $\theta_1,\theta_2$ of the two critic networks in $agent_i$ using Eq (24)
12: update the actor network parameters $\phi$ of $agent_i$ using Eq (26)
13: update the entropy coefficient $\alpha$ of $agent_i$ using Eq (27)
14: update the parameters $\bar{\theta}_1,\bar{\theta}_2$ of the two Target Network in $agent_i$ using Eq (23)
15: **If** $done$ **do**
16: **Break**
17: update $\mathrm{L}$ based on the final node position
18: extract the selected starting node $s'$ and routing path $\mathrm{L}'$ based on $\mathrm{L}$
19: calculate the feedback reward $R$ using Eq (13) based on $\mathrm{L}'$.
20: return $s'$, $\mathrm{L}'$, $R$
21: **END**

The algorithm takes the current network topology $\mathfrak{G}=(\mathcal{V},\mathcal{E})$, the network state information matrix $\mathrm{TM}$, the set of departure nodes $srcs$, the SAC agent (including its network parameters) $agent_i$, and the batch update size $bs$ as inputs. The outputs are the optimal departure node $s'$, selected from the current set of departure nodes for the subtask $srcs$ and the optimal routing path $\mathrm{L}'$ from this node to the destination node $v_{d_i}$. Line 1 initializes the round's state information and path list. Lines 3–7 represent the agent's single-step sampling to obtain data samples $(s_t, a_t^{sac}, r_t, s_{t+1}, done, mask_{s_t})$ and place them into the experience replay pool. Lines 8–16 constitute the agent's algorithm parameter update steps. Lines 17–20 calculate the path feedback reward for the upper layer at the end of the round.

The framework of the MA-DHRL-OM algorithm designed in this paper is shown in Algorithm 2.

**Algorithm 2**: MA-DHRL-OM

**Input:** Global network topology $\mathfrak{G}=(\mathcal{V},\mathcal{E})$, upper-layer PPO policy network learning rate $\alpha_a$, value network learning rate $\alpha_c$, total algorithm iterations $Episodes$, upper-layer PPO updates per round $eps$, lower-layer SAC multi-agent $agent_i, i=1,\ldots,n_{vd}$

**Output:** OM tree from source node $v_s$ to destination node $v_{d_i}$

1： calculate the matrix $NDM, SNDM$ based on the network topology: $\mathfrak{G}$

2： initialize the parameters $\theta, \phi$ of the policy network and value network in the initialization of the upper-layer PPO Algorithm

3： initialize the critic network parameters $\theta_1, \theta_2$, target network parameters $\overline{\theta}_1, \overline{\theta}_2$, and actor network parameters $\phi$ for all $agent_i$ in the lower-level SAC Algorithm

4： **For** $e=1 \rightarrow Episodes$ **do**

5： extract TM at a given time from the network state information database

6： initialize the state $s_t$ for the upper-layer PPO algorithm based on the source node $v_s$, the destination node set $\mathcal{V}_d$, and the matrix $NDM, SNDM$

7： **For** time step $t=1 \rightarrow \mathrm{T}$ **do:**

8： get $mask_{s_t}$ from environment based on State $s_t$

9： select $a_t = \pi_\theta(s_t)$ based on Policy $\theta$, State $s_t$, and $mask_{s_t}$

10： PPO agent executes the action $a_t$, the state changes to $s_{t+1}$ and return $done$

11： **If** $t=\mathrm{T}$ **do**

12： divide tasks using sequence C. Each agent executes Algorithm 1

13： feeds back reward signals aggregated to the upper layer, which receives the final reward $r$

14： **Else**

15： Retrieve intermediate rewards $r$ from the upper layer

16： Calculate $A_t$ using GAE with the sampling sequence tuple $\left\{\left(s_t, a_t^{PPO}, r_t, s_{t+1}, done, mask_{s_t}\right)\right\}_{t=1}^{T}$

17： **For** Training rounds $k=1 \rightarrow eps$ **do**

18： Calculate the new-old policy ratio $r(\theta)$ using Eq (14)

19： Calculate the policy network loss according to Eq (15)

20： Calculate the value network loss according to Eq (16)

21： Update policy network $\theta \leftarrow \theta - \alpha_a \nabla_\theta \mathcal{L}^{CLIP}$ and value network $\phi \leftarrow \phi - \alpha_c \nabla_\phi \mathcal{L}^{V}$

22： **END**

The algorithm inputs include the network topology, the learning rate of the upper-layer PPO algorithm ($\alpha$), the total number of iterations for the algorithm ($Episodes$), the number of updates per round for the PPO algorithm ($eps$), and the number of multi-agent systems for the lower-layer SAC algorithm. The output is the optimal cost-overlay multicast tree. Lines 1–3 initialize algorithm parameters and the environment. Lines 5 and 6 initialize the environmental state for each round and extract network state information. Lines 8–10 obtain the action output and next state for each step of the PPO agent. Lines 12–15 acquire rewards for the upper-layer PPO algorithm, with Line 13 obtaining the final sequence reward $R_{Seq}$. Multiple agents divide tasks sequentially, concurrently completing corresponding subtasks according to Algorithm 1 and feeding aggregated rewards to the upper-layer PPO algorithm. Lines 16–21 describe how the PPO agent updates network parameters based on sampling data tuples.

## 6. Experimental setup and performance evaluation

*6.1. Simulation environment design*

The experiments are conducted on the Mininet-WiFi 2.3.1b simulation platform [42], which supports the construction of highly customizable and complex Software-Defined Networking (SDN) environments. The hardware configuration of the experimental server is as follows: Intel(R) Xeon(R) Gold 5218 CPU @ 2.30 GHz and NVIDIA GeForce RTX 3090 GPU (24 GB). The software environment includes Ubuntu 18.04.2 LTS, Python 3.9, PyTorch 2.1.1, and CUDA version 11.8. The source code is developed remotely on the server using PyCharm 2021.

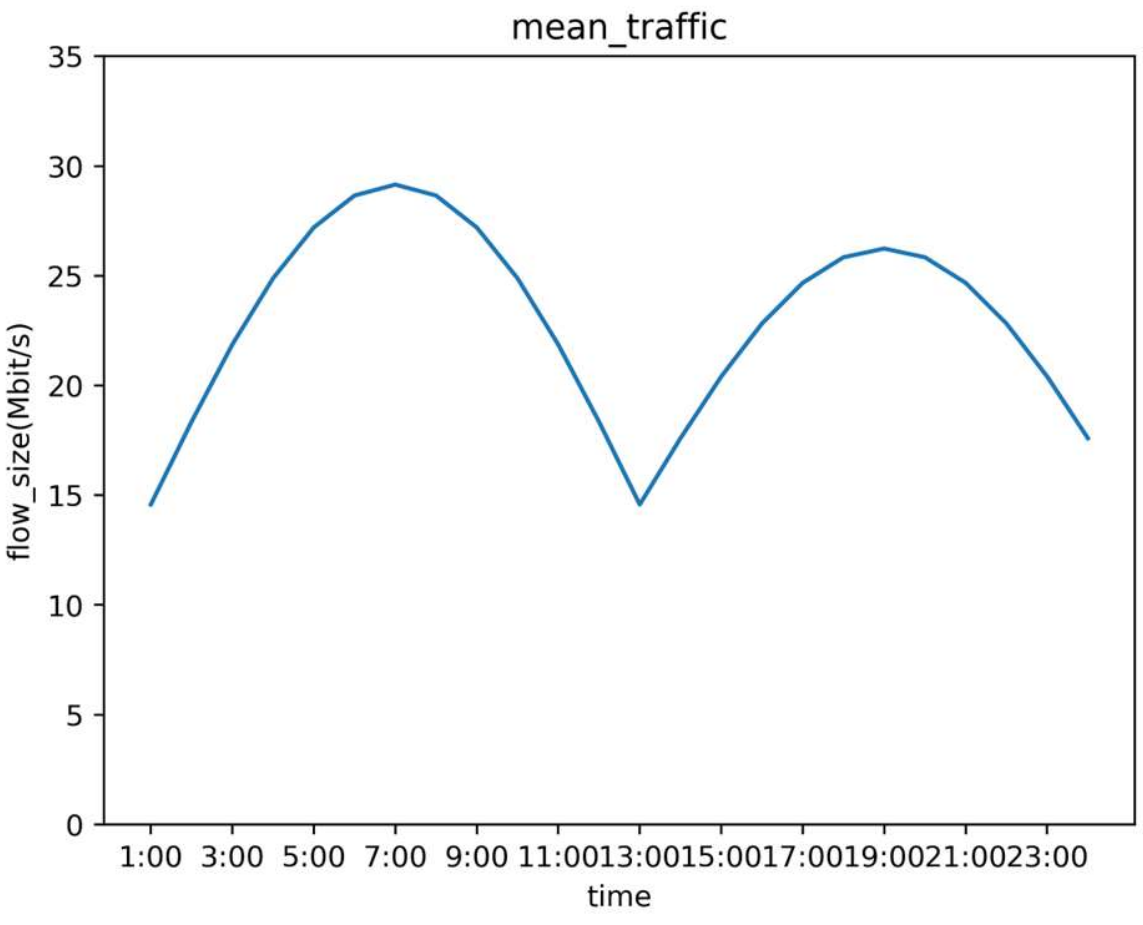

**Figure 4.** Traffic generation using Iperf.

To simulate realistic network traffic, the Iperf traffic generation tool [43] is used to transmit UDP (User Datagram Protocol) packets between nodes. Following the methodology in [30], we emulate the 24-hour variation of network traffic over a full day, as shown in Fig. 4. In Fig. 4, the horizontal axis represents time, while the vertical axis denotes the average sending rate per node in Mbit/s. The overall traffic trend conforms to the typical diurnal network load patterns observed at different times of the day.

The SDN controller used in this study is Ryu 4.3.4 [44], which is primarily responsible for handling events such as flow table installation for network nodes. During the experiments, Ryu collects network state information, converts it into graph-structured data, and stores it in Pickle format, thereby constructing the network traffic dataset.

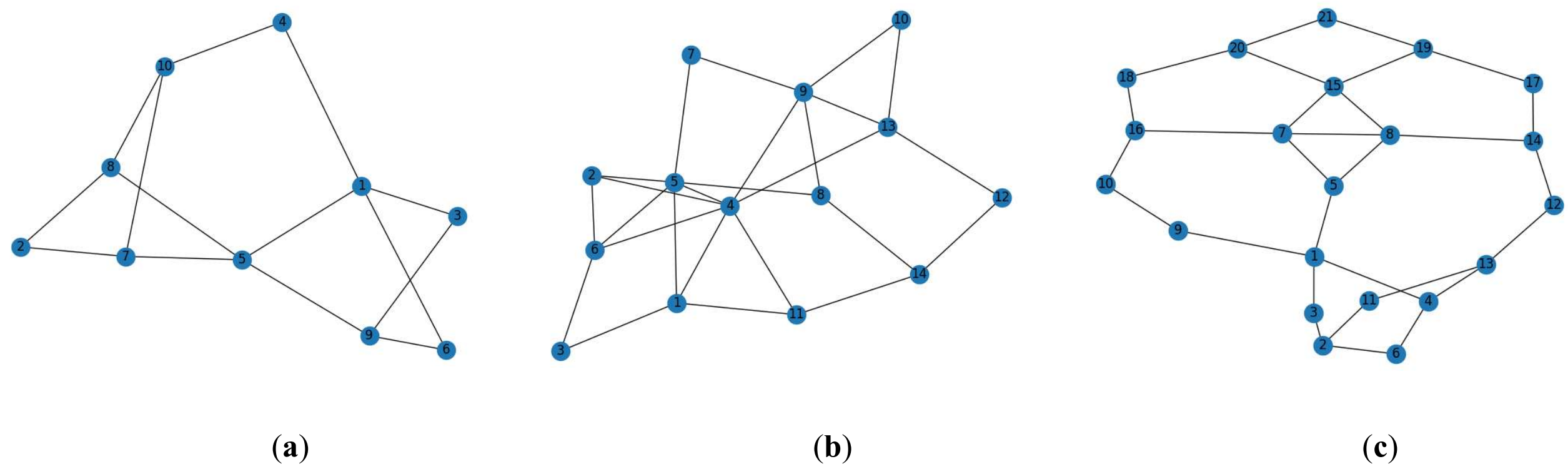

**Figure 5.** Three network topologies tested in the experiments: (**a**) 10NodeNet; (**b**) 14NodeNet; and (**c**) 21NodeNet.

Following the methodology in [45], three network topologies are adopted for performance evaluation, consisting of 10, 14, and 21 nodes, respectively, and named 10NodeNet, 14NodeNet, and 21NodeNet. The corresponding topologies are shown in Figure 5a–c. The link parameters in each topology are randomly generated according to a uniform distribution, where the link bandwidth ranges from 5 to 40 Mbps, and the link delay varies between 1 and 10 ms. The detailed configurations of the three experimental topologies and the corresponding test data are presented in Table 1.

**Table 1.** The configurations and test data of experimental topologies.

| Configurations and data | 10NodeNet | 14NodeNet | 21NodeNet |
|---|---|---|---|
| Number of nodes | 10 | 14 | 21 |
| Number of links | 14 | 24 | 29 |
| Number of destination nodes | 3 | 3 | 3 |
| Link bandwidth range (Mbps) | [5,40] | [5,40] | [5,40] |
| Link delay range (ms) | [1,10] | [1,10] | [1,10] |
| Link packet loss rate range (%) | [0,5] | [0,5] | [0,5] |
| State collection interval (s) | 5 | 5 | 5 |
| SDN ctrl max interact. interval (s) | 1.3 | 1.3 | 1.3 |
| Inference time (per iter, ms) [mean ± std] | 18.31 ± 1.33 | 21.56 ± 1.49 | 29.31 ± 2.68 |
| Training time (per iter, ms) [mean ± std] | 172.81 ± 8.58 | 196.36 ± 13.58 | 272.78 ± 15.75 |

As shown in the experimental results in Table 1, the per-step training time and inference time increase as the network size grows; however, their growth remains sublinear. This indicates that the proposed method maintains a certain degree of scalability in scenarios where the number of nodes continues to increase.

*6.2. Performance metrics*

We employ the bottleneck bandwidth, latency, and packet loss rate of the overlay multicast spanning tree as evaluation metrics for the proposed algorithm, as shown in Eq (28).

$$\begin{aligned} \overline{bw_T} &= average\frac{\sum_{p_k \in T} bw_k}{K} \\ \overline{delay_T} &= average\frac{\sum_{p_k \in T} delay_k}{K} \\ \overline{loss_T} &= average\frac{\sum_{p_k \in T} loss_k}{K} \end{aligned} \tag{28}$$

where $\overline{bw_T}$, $\overline{delay_T}$, and $\overline{loss_T}$ represent the average remaining bandwidth, average delay, and average packet loss rate of the overlay multicast tree, respectively; $bw_k$, $delay_k$, and $loss_k$ denote the bottleneck Bandwidth, latency, and packet loss rate, respectively, of the routing path $p_k$ reaching destination node $v_{d_k}$ in the tree; and $\mathrm{K} = n_{vd}$ is the number of destination nodes in the tree, which also corresponds to the number of unicast paths within it.

*6.3. Reinforcement learning parameter configuration*

Regarding the values of weight $\beta_i, i = 1,2,3$ mentioned in the path performance cost Eq (6) and the agent reward mechanism Eq (20), to avoid unbalanced decisions caused by excessive bias toward a single metric, we assign a higher weight to residual bandwidth due to its greater influence on improving network throughput and mitigating link congestion. Moreover, appropriate weights are retained for delay and packet loss rate, as real-time performance and reliability remain critical constraints for path selection to prevent unbalanced decisions from over-emphasizing a single metric. After extensive trials with different value combinations, the final weights for bandwidth, delay, and packet loss rate are set to 0.7, 0.2, and 0.1, respectively.

The value ranges and tuning ranges of the major parameters in the experiments of this paper are shown in Table 2.

The learning rate is a critical hyperparameter in deep reinforcement learning, determining the step size during each parameter update. An excessively high learning rate may cause training instability or prevent model convergence. Conversely, an overly low learning rate can lead to slow convergence, impairing learning efficiency and final performance. Here, we employ a hierarchical reinforcement learning algorithm framework integrating the Actor-Critic architecture from PPO and SAC. To systematically analyze the impact of learning rate on model performance, experiments fix the network learning rate in PPO and the actor learning rate in SAC while varying and comparing the learning rate of the critic network in SAC. Experimental results are shown in Figure 6.

Experimental results indicate that when $\mathrm{critic_lr} = 0.001$, a large learning rate makes reward convergence difficult. When $\mathrm{critic_lr} = 0.0002$, an excessively small learning rate prevents the network from learning effectively, and reward values fail to converge. When $\mathrm{critic_lr} = 0.0005 \text{ or } 0.0007$, reward values can converge, but $\mathrm{critic_lr} = 0.0007$ converges faster and more stably. At this point, the critic learning rate is fixed while the actor learning rate is adjusted and compared. Experimental results are shown in Figure 7.

**Table 2.** The value ranges and tuning ranges of the major parameters in the experiments.

| Parameters | Value range | Tuning range |
| --- | --- | --- |
| Number of training episodes | [200, 2000] | 1000 |
| SAC actor learning rate | [1e-5, 1e-3] | {0.00001, 0.00005, 0.0001, 0.0002} |
| SAC critic learning rate | [1e-4, 1e-2] | {0.0002, 0.0005, 0.0007, 0.001} |
| SAC batch size | [32, 256] | {64, 128, 192, 256} |
| SAC entropy coefficient learning rate | [1e-4, 1e-2] | 0.001 |
| SAC soft update parameter | [0.001, 0.1] | 0.05 |
| SAC replay buffer size | [$1\times10^4$, $5\times10^5$] | 100,000 |
| SAC discount factor | [0.85, 0.99] | 0.98 |
| PPO actor learning rate | [1e-4, 1e-2] | 0.001 |
| PPO critic learning rate | [1e-3, 1e-1] | 0.01 |
| PPO discount factor | [0.85, 0.99] | 0.90 |
| PPO GAE parameter | [0.90, 0.99] | 0.95 |
| PPO training epochs | [5, 20] | 10 |
| PPO clipping parameter | [0.1, 0.3] | 0.2 |

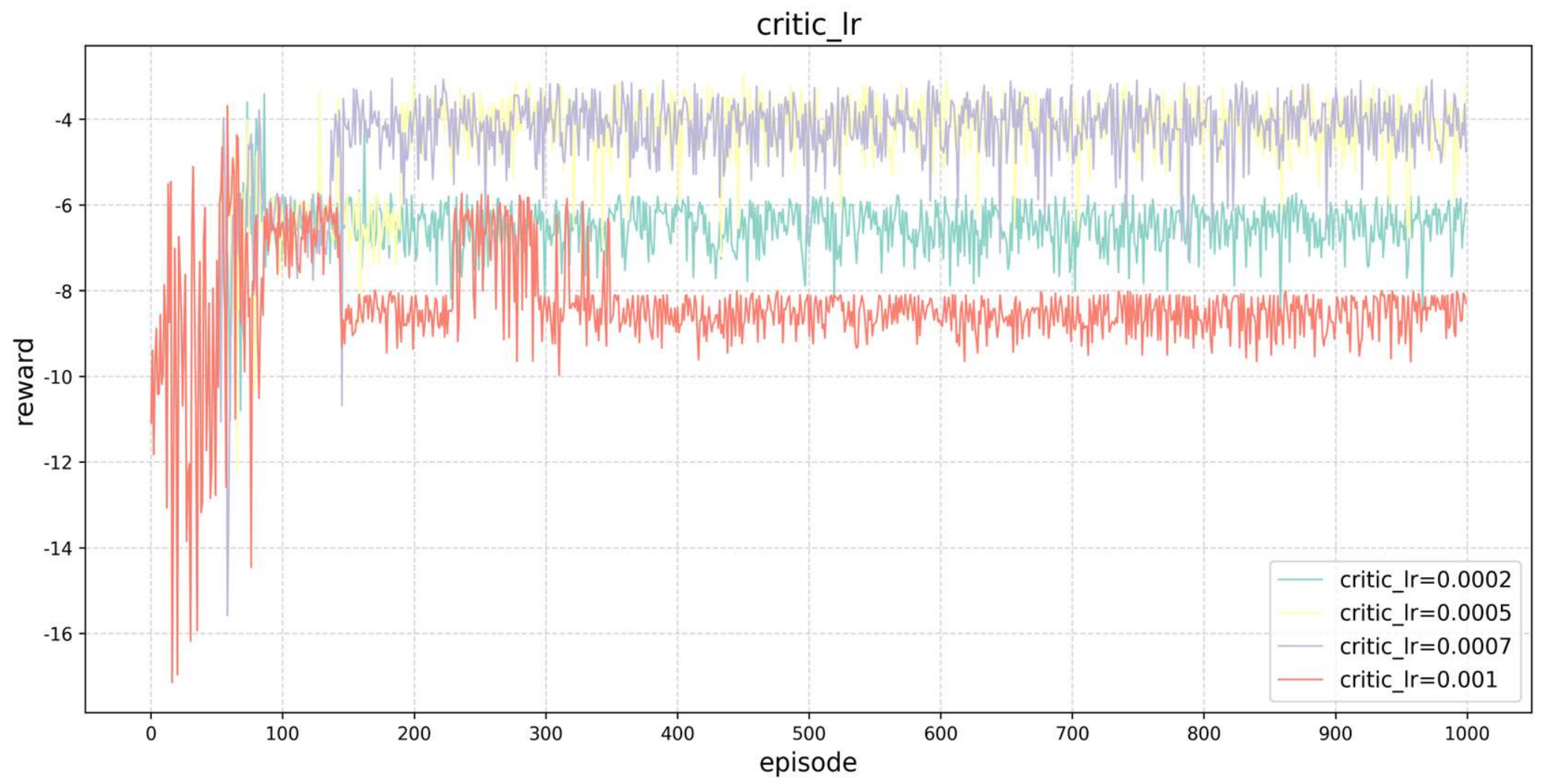

**Figure 6.** Critic learning rate comparison.

Experimental results indicate that increasing the learning rate to 0.00005 accelerates reward curve convergence. At actor_lr = 0.00001, the reward curve converges but at a slower pace. Raising the learning rate to 0.00005 further speeds convergence. At 0.0001, the reward curve converges most rapidly compared to other parameter settings.

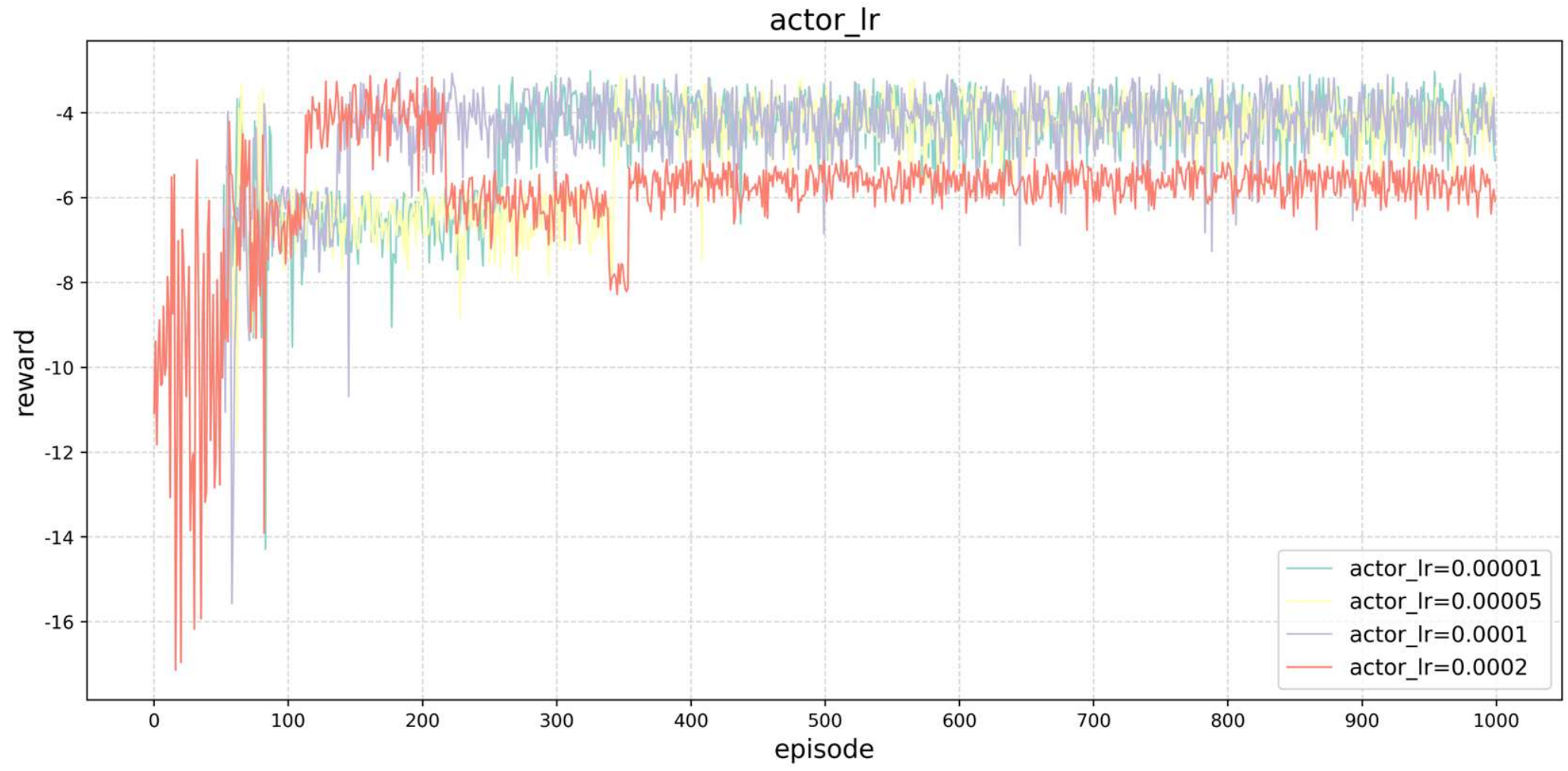

**Figure 7.** Actor learning rate comparison.

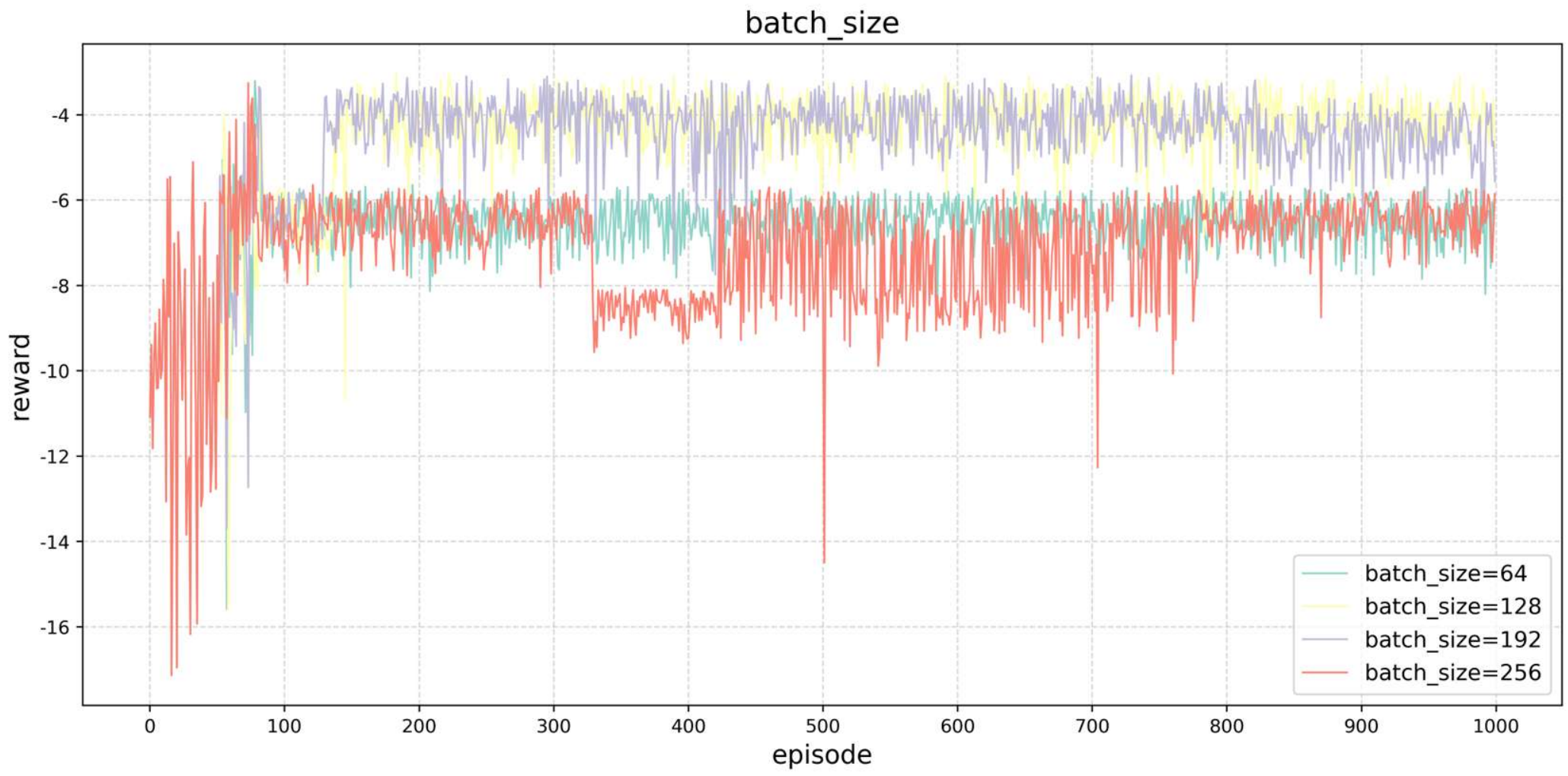

**Figure 8.** Batch size comparison.

Now, with all learning rates fixed, we adjust the batch size parameter for comparison. The experimental results are shown in Figure 8.

Experimental results indicate that when $\text{batch_size} = 64$, the variance of the algorithm's gradient estimation increases, causing training instability and difficulty in reward convergence. When $\text{batch_size} = 256$, the agent overfits to the local dataset, preventing reward convergence. When $\text{batch_size}$ is set to 128 or 192, reward convergence is achieved, with $\text{batch_size} = 128$ yielding the most stable convergence.

To more objectively evaluate the stability of the proposed method under random initialization, we conduct repeated experiments with multiple random seeds on the 14NodeNet topology. The key performance metrics are collected and summarized in Table 3.

**Table 3.** Key performance metrics under random seeds (N = 5).

| Random seed | Bottleneck bandwidth (Mbps) | Delay (ms) | Packet loss rate (%) |
|---|---|---|---|
| Seed = 1 | 20.42 | 7.22 | 0.0008 |
| Seed = 11 | 15.32 | 10.67 | 0.0010 |
| Seed = 21 | 18.82 | 6.22 | 0.0012 |
| Seed = 31 | 18.82 | 5.46 | 0.0017 |
| Seed = 41 | 20.42 | 7.21 | 0.0008 |
| Mean ± Std | 18.76 ± 2.08 | 7.36 ± 1.99 | 0.0011 ± 0.0004 |

The results in Table 3 show that the proposed method maintains relatively stable performance overall under multiple random seeds. In general, the bottleneck bandwidth remains around 15–20 Mbps, the end-to-end delay is usually kept within a low range, and the packet loss rate stays close to zero. This indicates that the proposed method is robust to random initialization and can achieve consistent routing performance across repeated runs.

*6.4. Comparative experiments*

To evaluate the performance of the MA-DHRL-OM algorithm, we simulate real-world traffic characteristics in wireless network topologies with 10, 14, and 21 nodes. The experimental setup designates one source node and three destination nodes to construct an overlay multicast communication scenario, thereby testing and analyzing the optimization performance of the resulting overlay multicast paths. The baseline algorithms include the traditional routing protocol OSPF and three single-agent learning methods: Actor-Critic, PPO, and SAC. Since overlay multicast is implemented over underlying unicast routes in this setting, the multicast problem is decomposed into building individual unicast paths from the source to each destination node for performance comparison. During evaluation, key performance metrics, including the bottleneck bandwidth, end-to-end latency, and packet loss rate of the unicast paths, are measured to comprehensively assess the performance of the MA-DHRL-OM algorithm in multi-destination transmission scenarios. All key experimental results are shown in Table 4, with values retained to 4 decimal places.

As shown in Figure 9a–c and Table 4, the evaluation metric is the average total link bottleneck bandwidth of the paths from the source node to each destination node for the agent. Experimental results show that for three topologies, 10NodeNet, 14NodeNet, and 21NodeNet, the MA-DHRL-OM algorithm improves the average path throughput by 9.73%, 19.09%, and 10.62%, respectively, compared with the traditional OSPF protocol; by 5%, 14.07%, and 5.97%, respectively, compared with the Actor-Critic algorithm; and by 6.61%, 15.7%, and 31.12%, respectively, compared with the PPO algorithm. The improvements over the SAC algorithm are 3.39%, 17.28%, and 13.77% for 10NodeNet, 14NodeNet, and 21NodeNet, respectively.

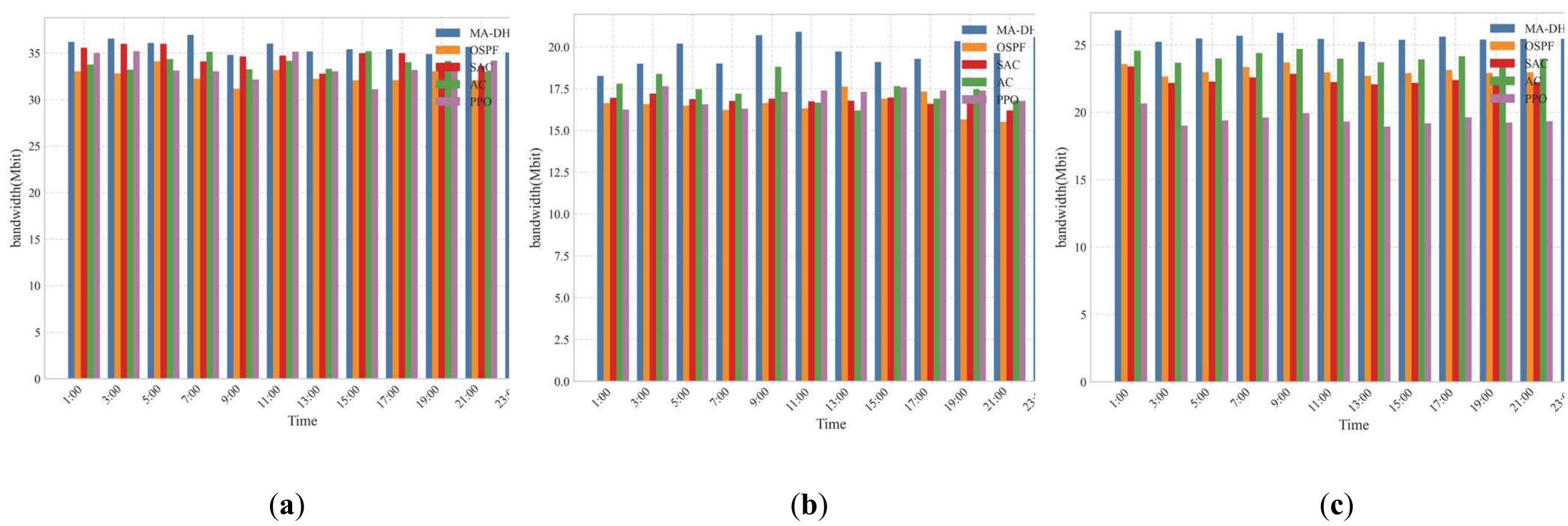

(**a**) (**b**) (**c**)

**Figure 9.** Bottleneck bandwidth experimental results for three network topologies: (**a**) 10NodeNet; (**b**) 14NodeNet; and (**c**) 21NodeNet.

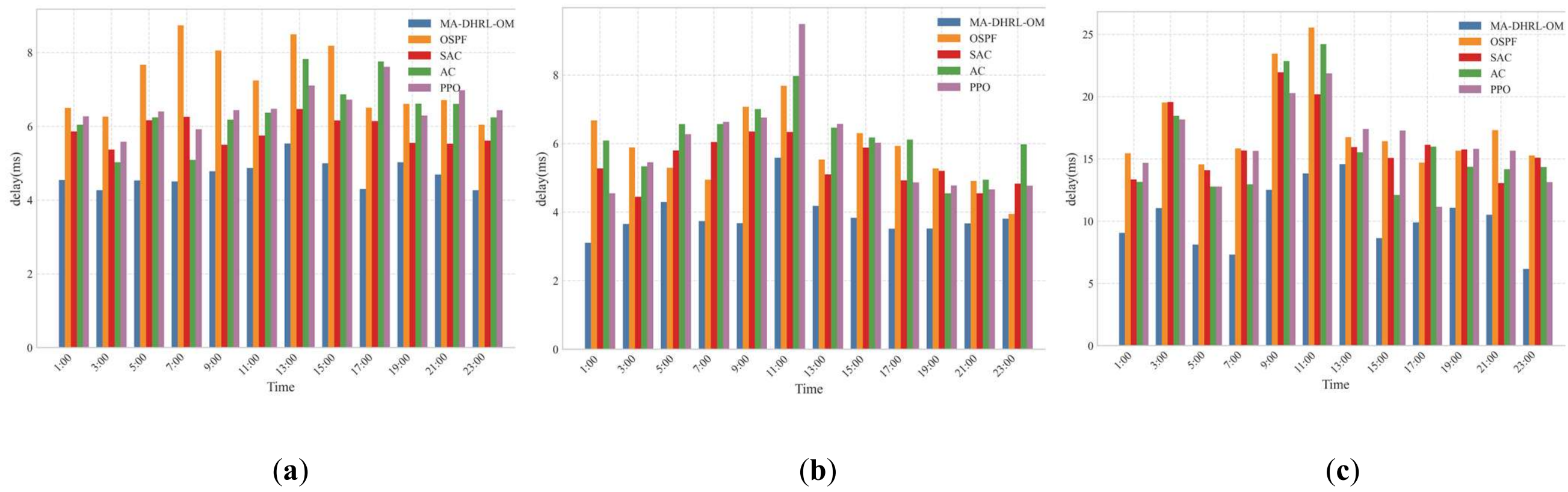

(**a**) (**b**) (**c**)

**Figure 10.** Average total link delay experimental results for three network topologies: (**a**) 10NodeNet; (**b**) 14NodeNet; and (**c**) 21NodeNet.

As shown in Figure 10a–c and Table 4, the evaluation metric is the average total link delay of the paths from the source node to each destination node for the agent. Experimental results indicate that in the three topologies, 10NodeNet, 14NodeNet, and 21NodeNet, the MA-DHRL-OM algorithm reduces the average delay by 7.17%, 49.07%, and 28.64%, respectively, compared with OSPF; by 36.5%, 58.31%, and 55.54%, respectively, compared with Actor-Critic; and by 38.94%, 52.03%, and 61.16%, respectively, compared with PPO. The reductions over the SAC algorithm are 3.28%, 7.66%, and 23.87% for 10NodeNet, 14NodeNet, and 21NodeNet, respectively.

Taken together, these results indicate that MA-DHRL-OM preferentially selects low-latency paths during route decision-making, thereby significantly reducing end-to-end transmission delay and better satisfying the performance requirements of latency-sensitive network applications.

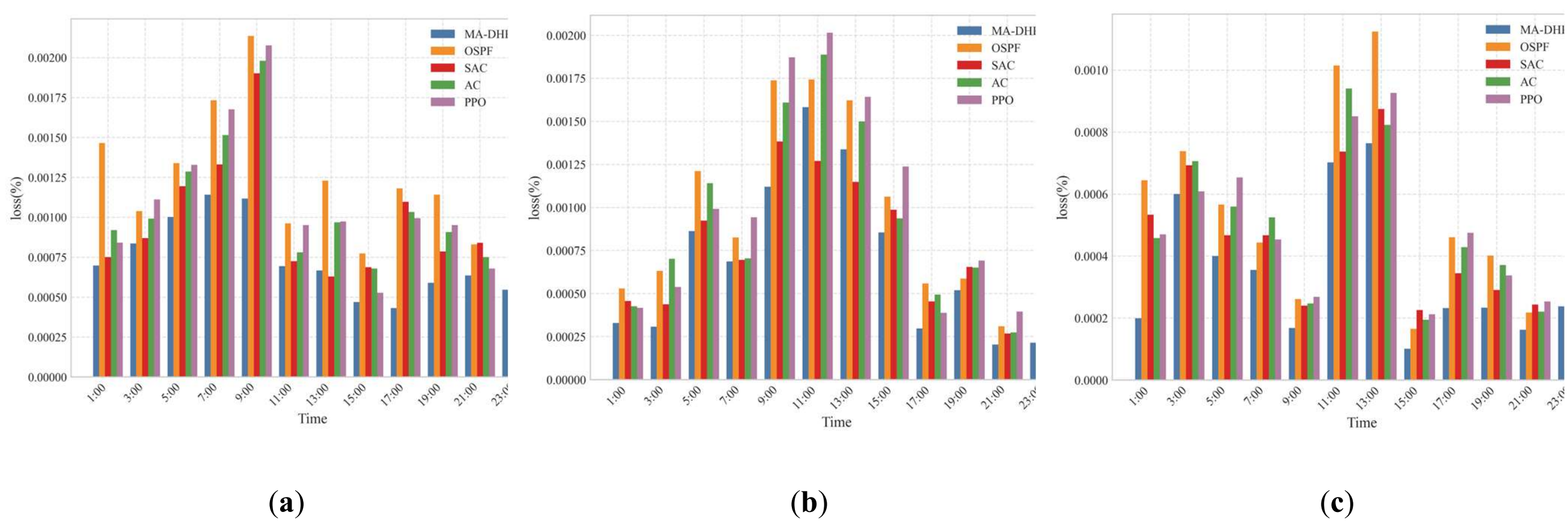

**Figure 11.** Average packet loss rate experimental results for three network topologies: (**a**) 10NodeNet; (**b**) 14NodeNet; and (**c**) 21NodeNet.

**Table 4.** Summary of key experimental results.

| Topo | Algorithm | Bottleneck bandwidth (Mbps) | Delay (ms) | Packet loss rate (%) |
|---|---|---|---|---|
| 10NodeNet | MA-DHRL-OM | 35.6919 | 3.8835 | 0.0007 |
| | OSPF | 32.5247 | 5.7892 | 0.0012 |
| | Actor-Critic | 33.9907 | 6.1484 | 0.0010 |
| | PPO | 33.4765 | 5.9044 | 0.0011 |
| | SAC | 34.5186 | 5.3966 | 0.0010 |
| 14NodeNet | MA-DHRL-OM | 19.7318 | 3.8835 | 0.0007 |
| | OSPF | 16.5681 | 5.7892 | 0.0009 |
| | Actor-Critic | 17.2973 | 6.1484 | 0.0009 |
| | PPO | 17.0532 | 5.9044 | 0.0010 |
| | SAC | 16.8233 | 5.3966 | 0.0008 |
| 21NodeNet | MA-DHRL-OM | 25.5281 | 10.2317 | 0.0003 |
| | OSPF | 23.076 | 17.5434 | 0.0005 |
| | Actor-Critic | 24.0894 | 15.9150 | 0.0005 |
| | PPO | 19.4685 | 16.1570 | 0.0005 |
| | SAC | 22.4382 | 16.3278 | 0.0005 |

As shown in Figure 11a–c and Table 4, the evaluation metric is the average packet loss rate of the paths from the source node to each destination node for the agent. Experimental results demonstrate that in the three topology structures, 10NodeNet, 14NodeNet, and 21NodeNet, the MA-DHRL-OM algorithm achieves performance improvements of 68.25%, 35.64%, and 57.58%, respectively, compared with OSPF; of 42.18%, 23.85%, and 42.12%, respectively, compared with Actor-Critic; and of 46.72%, 37.66%, and 49.56%, respectively, compared with PPO. The improvements over the SAC algorithm are 31.97%, 9.72%, and 30.31% for 10NodeNet, 14NodeNet, and 21NodeNet, respectively. Given that network links are prone to high packet loss under heavy load or congestion, MA-DHRL-OM effectively avoids congested links through its path optimization strategy. Consequently, it significantly reduces the average packet loss rate across topologies, thereby substantially enhancing the overall stability and reliability of data transmission.

## 7. Conclusions

In this paper, we propose a multi-agent hierarchical reinforcement learning-based OM routing method. First, addressing the limitations of traditional OM methods, such as insufficient awareness of underlying network state and difficulty adapting to dynamic link changes, MA-DHRL-OM incorporates an SDN architecture. This leverages the controller's global perspective and programmability to achieve real-time link state awareness and intelligently adjust OM paths dynamically. Second, addressing the high-dimensional NP-hard decision problem inherent in overlay multicast, MA-DHRL-OM employs a multi-agent and hierarchical reinforcement learning framework. It divides the task into two phases: Generating a sequence of destination nodes and intelligently selecting the route path from the source node. These phases are collaboratively executed by upper- and lower-layer agents to construct the overlay multicast tree.

Experimental results demonstrate that under network topologies, MA-DHRL-OM significantly outperforms traditional routing protocols and mainstream reinforcement learning algorithms in key performance metrics such as path throughput, latency control, and packet loss rate. MA-DHRL-OM exhibits higher stability and generalization capabilities, fully validating its application potential and practical value in dynamic complex network environments.

In future work, we will further advance overlay multicast routing strategies under the SDN architecture, with particular emphasis on the impact of dynamic member joins and leaves on path stability and performance variability. From an algorithmic perspective, computational overhead will be systematically evaluated under a training–inference separation framework, quantifying convergence time and online inference latency across topology scales and traffic intensities. Resource consumption and control-plane load after deployment within the SDN controller will also be analyzed to assess practical feasibility. The study will be extended to larger-scale topologies to examine how increasing numbers of nodes and destinations affect decision dimensionality, convergence behavior, and overall operational efficiency, thereby validating the scalability and stability of the hierarchical and multi-agent mechanisms. In addition, a dynamic stress-testing environment will be established to investigate path reconstruction overhead and policy reconvergence under frequent membership changes and link-state perturbations, enhancing the robustness and practical applicability of the proposed approach in complex dynamic networks.

## Use of AI tools declaration

The authors declare they have not used Artificial Intelligence (AI) tools in the creation of this article.

## Acknowledgments

This work was supported by the National Natural Science Foundation of China (No. 62161006), Guangxi Science and Technology Program under Grant (No. FN2504240022), and the Subsidization of Innovation Project of Guangxi Graduate Education (No. YCSW2025351).

## Conflict of interest

All authors declare no conflicts of interest in this paper.